\documentclass[range]{ar2e}
\usepackage{graphicx}

\newcommand{\beq}{\begin{equation}}
\newcommand{\eeq}{\end{equation}}
\newcommand{\bea}{\begin{eqnarray}}
\newcommand{\eea}{\end{eqnarray}}

\newcommand{\nn}{\nonumber}
\newcommand{\benn}{\begin{displaymath}}
\newcommand{\eenn}{\end{displaymath}}

\begin{document}

\title{ Time-Dependent Density Functional Theory and the Real-Time Dynamics of Fermi Superfluids}

\author{Aurel Bulgac \affiliation{Department of Physics, University of Washington, Seattle, WA 98195--1560, USA \\ Email: \quad bulgac@uw.edu}}
 
\markboth{Real-Time Dynamics of Fermi Superfluids}{ Real-Time Dynamics of Fermi Superfluids}

 
\begin{abstract}

I describe the Time-Dependent Superfluid Local Density Approximation, which is an adiabatic extension of the Density Functional Theory to superfluid Fermi systems and their real-time dynamics.  This new theoretical framework  has been applied to describe a number of phenomena in cold atomic gases and nuclear collective motion: excitation of the Higgs modes in strongly interacting Fermi superfluids, generation of quantized vortices, crossing and reconnection of vortices, excitation of the superflow at velocities above the critical velocity, excitation of quantum shock waves and domain walls in the collisions of superfluid atomic clouds, excitation of collective states in nuclei. 

\end{abstract}

\date{\today}
 
\maketitle

-------------------------------------------------

\section{Density Functional Theory}

In a remarkable theorem proven almost fifty years ago, Kohn and collaborators \cite{hk,ks,wk} established that there is an one-to-one map between the ground state wave function of an interacting multi-electron system, the number density, and the external Coulomb one-body potential created by the (static) nuclei. Even though the theorem was formulated in terms of electrons and Coulomb interaction among them and with nuclei, the proof never relies on these specific aspects and it applies to any non-relativistic fermion system and as such it is widely referred to as the Density Functional Theory (DFT).  Subsequently this theorem has been extended to apply to more general situations \cite{monograph1,monograph3} and since then it has became a workhorse in chemistry and condensed matter studies, due to the tremendous mathematical simplification achieved by replacing the many-body Schr\"{o}dinger equation with a system of non-linear and coupled 3D partial differential equations, formally equivalent to a meanfield treatment of the electron systems. The main difficulty in applying DFT resides in the fact that the generation of the corresponding Energy Density Functional (EDF) is more of an art than a science, as the Hohenberg-Kohn \cite{hk} and Kohn-Sham \cite{ks} theorems do not provide for an algorithm to derive the EDF from the many-body Schr\"{o}dinger equation. One very important generalization of the original DFT approach was the development of the time-dependent version of DFT \cite{tddft,monograph4,monograph2}, which allows in principle the replacement of the time-dependent many-body Schr\"{o}dinger equation with a set of time-dependent coupled 3D partial differential equations, formally equivalent to a time-dependent meanfield approach. Thus ``Time-dependent density functional theory (TDDFT) can be viewed as an exact reformulation of time-dependent quantum mechanics, where the fundamental variable is no longer the many-body wave-function but the density." \cite{org}. Since its formulation TDDFT has been applied mainly in atomic and molecular calculations for the study of the excited states of many-electron systems and by now it has achieved a very sophisticated level of development as amply highlighted in recent monographs \cite{monograph4,monograph2}. While in principle DFT as initially formulated can be used to describe any interacting fermion system, the application to superfluid fermionic systems is clearly going to encounter some challenges. If one were to use only the density of the many fermion system it would be impossible to distinguish between a superfluid and a normal system. One can easily envision a situation when a superfluid system is stirred energetically enough that in some regions or even globally the system can undergo a transition to a normal state. The need to develop a version of the DFT suitable for the description of superfluid systems was recognized quite some time ago by Oliveira, Gross and Kohn \cite{oliveira}. The version of DFT suggested by these authors however lacked the great simplicity of the Local Density Approximation (LDA) of Kohn and Sham \cite{ks}, namely of being formally equivalent to a local implementation of the meanfield approximation. Similarly to the approach of Bardeen, Cooper and Schrieffer  \cite{bcs} Gross {\it et al} \cite{oliveira,kurth,profeta} introduced the anomalous density to describe the order parameter in the superfluid phase. The anomalous density however has an ultraviolet divergence if the pairing field is a local potential \cite{ab88,castin,by,ab02} and a meaningful theoretical framework can be developed only if one introduces and appropriate regularization and renormalization procedures, similar to those routinely used in quantum field theories. Gross {\it et al} \cite{oliveira,kurth,profeta} eschewed these issues by considering a non-local version of the DFT in the case of superfluid systems, with a non-local pairing field.  

 In the case of Bose superfluids,  ``two phenomenological theories explain almost all experiments to date" \cite{PhysicsToday}, the two-fluid hydrodynamics  and the ``complementary view, provided by Fritz London, Lars Onsager, and Richard Feynman, (that) treats the superlfuid as a macroscopic quantum state" \cite{PhysicsToday}.  Landau \cite{landau} identified the excitations which he called rotons with motion characterized by non-vanishing superfluid velocity circulation ($\vec{\nabla}\times \vec{v} \neq 0$) and he did not envision the existence of quantized vortices. When a superfluid is brought into rotation, quantum vortices, predicted by Onsager and Feynman \cite{onsager,feynman},  are formed and the two-fluid hydrodynamics is unable to describe their dynamic generation. Even though Landau described his approach initially as a quantum theory \cite{landau},  only in its classical incarnation has ever been used in practice, and the two-fluid hydrodynamics \cite{tisza1,tisza2,landau,khalatnikov} is ``essentially thermodynamics" \cite{PhysicsToday}. It is ultimately a phenomenological classical approach in which Planck's constant never explicitly enters. Later on, vortex quantization was imposed by hand when needed \cite{khalatnikov}, in a manner similar to the Bohr quantization rule of the hydrogen atom, following Onsager's and Feyman's quantization conditions \cite{onsager,feynman}.  
 
 In the case of dilute Bose systems the Gross-Pitaevskii \cite{gross,pitaevskii} equation can be derived 
\begin{equation}
i\hbar\dot{\Psi}(\vec{r}, t) = -\frac{\hbar^2\Delta}{2m}\Psi(\vec{r}, t) + g|\Psi(\vec{r}, t)|^2\Psi(\vec{r}, t) + V_{ext}(\vec{r}, t)\Psi(\vec{r}, t), \label{eq:GP}
\end{equation}
where $g>0$ is a coupling constant for the boson-boson interaction and $V_{ext}(\vec{r}, t)$ is an external potential. This equation can be used to describe the real-time evolution of a bosonic superfluid at very low temperatures, and in particular the vortex generation. A similar description did not exist for a fermionic superfluid until recently.  Many phenomena in cold atom physics, nuclear physics and neutron star crust  require a real-time dynamical approach, often beyond the linear response regime and for time scales when the role of collisions can be neglected. The description of such phenomena lead  the need for a theoretical framework of the real-time dynamics of a fermion superfluid arises. 
 
 A local extension of DFT to superfluid systems in the spirit of Kohn-Sham LDA has been formulated  in Refs. \cite{by,ab02,bfm} and was dubbed the Superfluid Local Density Approximation (SLDA) as being a natural extension of the Kohn-Sham LDA to superfluid systems.  SLDA and its time-dependent extension TDSLDA was applied to a range of static and time-dependent  situations in nuclear and cold atom systems \cite{bfm,slda,vortex,bf,higgs,vortices,qsw,yb,nvortex,ionel}. In its present formulation SLDA has been implemented only for phenomena at zero temperature, and its extension to time-dependent processes is valid only as long as collisions, leading to entropy production and thermalization of the system can be ignored. In this respect TDSLDA is similar in spirit to the Landau's Fermi liquid theory, which describes the propagation of the high-frequency zero-sound at not very long time scales, as opposed to a kinetic approach necessary to describe the propagation of the relatively lower frequency first-sound at much longer time scales. In the case of the zero-sound the local Fermi distribution does not equilibrate, while the opposite is true in the case of the first-sound. The high-frequency and low-frequency are determined when compared to the local relaxation rate, governed by the collision integral in a Boltzmann description of such  system. In the small amplitude limit the TDSLDA equations in the frequency representation become formally equivalent to the Landau's Fermi liquid theory or the linear response theory in the presence  of pairing correlations (or their absence, if the pairing field vanishes). 
 
\section{Unitary Fermi Gas}

In 1999 Bertsch introduced a  hypothetical system which later became known as the Unitary Fermi Gas (UFG) \cite{bertsch}. At the time this was a pure theoretical model of very dilute neutron matter, which subsequently became an object of intense study both theoretically and experimentally in the cold atom physics \cite{giorgini,bloch,ketterle, grimm,luo,bcsbec}. The UFG has properties surprisingly close to those of realistic dilute neutron matter \cite{gezerlis}. A UFG is a system of spin-1/2 fermions interacting only in the s-wave with an infinite scattering length and a zero effective range. The only Trivial dimensional arguments show that the energy of a homogeneous UFG is a function of only the Planck's constant $\hbar$, the fermion mass $m$, the volume of the system $V$ and the number of fermions $N$. The only quantity with dimension of energy one can form out of these constants  is $\xi\times 3\varepsilon_FN/5$, where $\xi$ is a dimensionless constant now called the Bertsch parameter, $\varepsilon_F = \hbar^2k_F^2/2 m$ is the Fermi energy of a free Fermi gas of the same density, and  $N/V = k_F^3/3\pi^2$. Since such an interaction is attractive $\xi<1$, but in 1999 it was not clear whether also $\xi>0$.  For $\xi=1$ this is the energy of a free Fermi gas, and if $\xi<0$ the system would collapse. Accurate quantum Monte Carlo (QMC) calculations $\xi=0.372(5)$ \cite{carlson1,carlson2,jc,forbes} and experimental measurements $\xi = 0.376(4)$ \cite{ku} have converged to basically the same value. Consequently a UFG is a gas, but somewhat surprisingly is also a superfluid, with one of the largest relative known pairing gaps in any fermion system $\Delta \approx 0.5 \varepsilon_F$. Superfludity in a UFG has been confirmed experimentally \cite{zwierlein} by directly putting in evidence the formation of the Abrikosov lattice of quantum vortices  \cite{abrikosov},  when such a system is brought into rotation by stirring it with laser beams in an atomic trap. 

The (TD)SLDA equations of motion are derived using an appropriately defined action integral, which is a straightforward generalization of the Kohn-Sham approach \cite{ks} to superfluid systems and time-dependent phenomena. This kind of extension of the Kohn-Sham approach to time-dependent phenomena is appropriately referred to as the Adiabatic Local Density Approximation (ALDA). If the number of spin-up and spin-down particles are equal, there is no spin-orbit interaction or velocity coupling, and if spin degrees of freedom are not excited, this action integral has the form:
\bea
\label{eq:tdslda}
&&\mathcal{S} = i\hbar\int dt d^3r \sum_{n}\left\{  u^*_{n}(\vec{r},t)\partial_t u_{n}(\vec{r},t)  +v^*_{n}(\vec{r},t)\partial_t  v_{n}(\vec{r},t)\right\} \nn \\ 
&&+\int dtd^3r \Biggl\{  \mathcal{E}\left [ n(\vec{r},t),\tau(\vec{r},t),\vec{j}(\vec{r},t),\nu(\vec{r},t)\right ]  + U(\vec{r},t) n(\vec{r},t)\Biggr\},
\eea
where $u_{n}(\vec{r},t)$ and $v_{n}(\vec{r},t)$ are quasi-particle wave functions (qspwf(s)),  and $\mathcal{E}$ is the energy density, which depends on  the number $n(\vec{r},t)=2\sum_n|v_{n}(\vec{r},t)|^2$, kinetic  $\tau(\vec{r},t)=2\sum_n|\vec{\nabla}v_{n}(\vec{r},t)|^2$,  current $\vec{j}(\vec{r},t)=2\,\mathrm{Im}[i\hbar \sum_n v_{n}^*(\vec{r},t)\vec{\nabla}v_{n}(\vec{r},t)] $, and anomalous density $\nu(\vec{r},t)=\sum_nv^*_{n}(\vec{r},t)u_{n}(\vec{r},t)$ respectively. $U(\vec{r},t)$ is some external potential in which the system might reside. Sometimes it is convenient to consider as well and external pairing field acting on the system, by adding to the above action integral a term $\Delta_{ext}(\vec{r},t) \nu^*(\vec{r},t) +  \Delta^*_{ext}(\vec{r},t) \nu(\vec{r},t)$. 

The energy density $\mathcal{E}$, as usual in DFT, is universal, in the sense that its form is independent of the external potential $U(\vec{r},t)$. This is the reason why  $\mathcal{E}$ should satisfy general symmetry principles: translational and rotational invariance, parity, local gauge and local Galilean invariance/covarince, and the theory should be renormalizable as well. As was mentioned above, in the case of superfluid systems the anomalous density and the kinetic energy density are both ultraviolet divergent quantities \cite{ab88,castin,by,ab02} and in order to remove these divergencies from the formalism both the kinetic and anomalous densities should enter the formalism in a unique combination. The UFG is a quire remarkable physical system, as one can use also simple dimensional arguments in conjunction with the symmetry requirements and renormalizability of the theory to show that the energy density has a very simple and essentially unique expression (if simplicity of the theory is invoked):
\bea\label{eq:edf}
\mathcal{E}\left [ n(\vec{r},t),\tau(\vec{r},t),\vec{j}(\vec{r},t),\nu(\vec{r},t)\right ]  =
\alpha \left[\frac{\hbar^2}{2m}\tau(\vec{r},t) -\Delta(\vec{r},t)\nu^*(\vec{r},t)\right ] + \nn \\
\beta \frac{3(3\pi^2)^{2/3}\hbar^2}{10m}n^{5/3}(\vec{r},t)+\lambda\frac{\hbar^2|\vec{\nabla}n(\vec{r},t)|^2}{m \, n(\vec{r},t)} +(1-\alpha)\frac{|\vec{j}(\vec{r},t)|^2}{m \, n(\vec{r},t)}.
\eea
The first term is the unique combination of the kinetic and anomalous densities required by the renormalizability of the theory; the second term is the only function of the density alone allowed by dimensional arguments in the case of a UFG (since no other dimensionfull parameters are needed to describe this system); the third term is lowest gradient correction and its amplitude is likely small \cite{gradient}; and the fourth term  is required by Galilean invariance \cite{slda}. The last term vanishes in the ground state where there are no currents, and the gradient correction term is absent in homogeneous matter. The constants $\alpha, \beta, \lambda$ and $\gamma$ (which is not shown here, see Refs. \cite{bfm,slda} for details, but enters in the definition of the pairing field $\Delta(\vec{r},t)$ \cite{slda}) are all dimensionless and they fully determine the energy density functional for a UFG. The constants  $\alpha, \beta, \gamma$ are determined by requiring that this energy density reproduce exactly the energy per particle, paring gap and effective mass obtained in accurate QMC calculations for a uniform system \cite{ab02,carlson1,carlson2}. The constant $\lambda$ is determined from reproducing the QMC results of an inhomogeneous  UFG in an external  trap \cite{gradient}. The magnitude of the gradient corrections is relatively small $\lambda \approx  - 0.1$ (and somewhat surprisingly negative) \cite{gradient}  and the effective mass is close to the bare mass value, as  $\alpha \approx 1.1$ \cite{bfm,slda,gradient}. Consequently, the last two terms in Equation (\ref{eq:edf}) represent somewhat small corrections when compared to the first two terms. Using this EDF (without gradient corrections, which were introduced later to further improve the agreement) one can predict now with impressive accuracy \cite{bfm,slda}  the results of the QMC calculations of inhomogeneous systems \cite{blume1,blume2} without any additional fitting and the agreement is always within the QMC errors. Even though more complicated EDF could be contemplated, which however will lack the simplicity of Equation (\ref{eq:edf}), there does not seem to be any need for them yet at the level of accuracy of the present QMC calculations. The absence of any dimensionfull scales, apart from the average inter-particle distance  $n^{-1/3}\approx \pi/k_F$, allows the determination of  the structure of the EDF for a UFG  essentially uniquely, and sets this strongly interacting system apart from other many-body systems. 

The equations for the qpwfs $u_{n}(\vec{r},t),v_{n}(\vec{r},t)$ have the time-dependent Bogoliubov-de Gennes \cite{degennes} form by design:
\begin{equation}\label{eq:tddft} 
\!\!\!   \left ( \begin{array}{c}   
    i \hbar \partial_t u_n(\vec{r},t)\\ 
   i \hbar \partial_t v_n(\vec{r},t) 
  \end{array}  \right )= 
  \left ( \begin{array}{cc}
    h(\vec{r},t)+U(\vec{r},t)  & \Delta(\vec{r},t) \\
    \Delta^*(\vec{r},t)& -h^*(\vec{r},t)-U(\vec{r},t) 
  \end{array}\right )
   \left ( \begin{array}{c} 
    u_n(\vec{r},t)\\      
    v_n(\vec{r},t)  \end{array} \right ).
\end{equation}
The operators here are defined as corresponding variational derivatives as usual: $h(\vec{r},t) = \delta \mathcal{E}/\delta n(\vec{r},t)$ and  $\Delta(\vec{r},t) = \delta \mathcal{E}/\delta \nu^*(\vec{r},t)$.  The fact that the pairing potential $\Delta(\vec{r},t)$ is a local operator, and not an integral one, is the reason why a Kohn-Sham type of implementation of DFT for superfluid systems displays ultraviolet divergencies in the kinetic and anomalous densities \cite{ab88,castin,by,ab02}, which require the implementation of a regularization and renormalization procedures described in Refs. \cite{by,ab02,bfm,slda}.  Numerically, the solutions of these equations is a formidable problem, since one has to solve an infinite system of time-dependent 3D nonlinear coupled partial differential equations, and this is feasible only on modern supercomputers \cite{roche}.   

\section{Nuclear Systems} 

During the last decade it was realized that the traditional nuclear meanfield-like approaches used for decades are nothing else but a disguised form  of DFT and the corresponding terminology has entered the theoretical nuclear physics field as well. The construction of an accurate nuclear EDF is an ongoing endeavor and several approaches are used. A number of nuclear theorists hope to handle the strong nuclear interactions within a many-body perturbation theory, using the modern chiral perturbation theory nucleon interactions to construct the ground state energy of nuclei \cite{dick1,dick2,dick3}, from which one can derive accurate enough local nuclear EDFs. Typically, however, various authors prefer a more phenomenological approach, using local EDFs which depend on proton and neutron number densities, spin and kinetic energy densities (and some of their derivatives), corresponding current densities, and proton and neutron anomalous densities as well \cite{bender,nmasses,witek,gfb,goriely}.  These nuclear EDFs  should satisfy the usual symmetries: translational and rotational symmetry; isospin symmetry which is broken explicitly by proton-neutron mass difference, Coulomb interaction and charge symmetry breaking forces (this one being routinely neglected in practice); parity, gauge, and Galilean invariance/covariance, and renormalizabilty (often treated in a rather naive manner).  Unfortunately, a number of these general requirements are often sacrificed in practice in the name of reaching agreement with experiment, see Ref. \cite{long} for a discussion of some these issues and of potential ways to further develop TDDFT. In nuclear systems there is a strong spin-orbit coupling, and both neutron and proton systems can become superfluid. In principle it is also possible that proton-neutron Cooper pairs are formed in some instances, though the experimental evidence is ambiguous at this time. It is possible as well that neutrons form pairs in $p-$ and $f$-waves (coupled by spin-orbit interaction) as well at higher densities in the neutron stars and the corresponding pairing field has a rather complicated spin-orbital structure. However, the basic difference of nuclei from  UFG is the the presence of two types of particles (protons and neutrons), spin-orbit coupling and  consequently the dependence of the EDF on several kinds of densities. However,  formally the corresponding equations for the qpwfs have a similar structure.

\section{Excitation of various collective modes}


\subsection{The Higgs mode}

\begin{figure}
\includegraphics[width=0.75\textwidth]{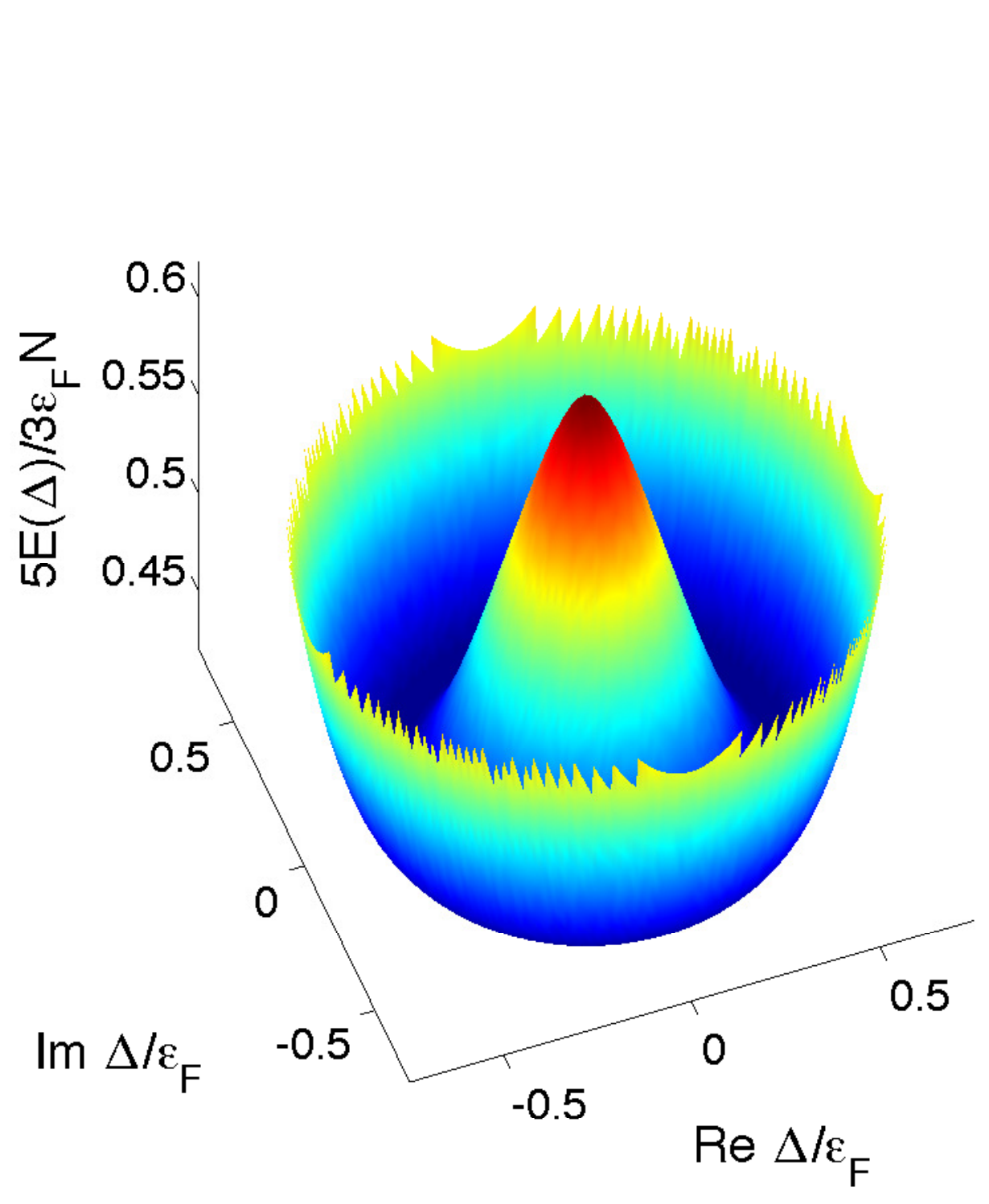}
\caption{The energy of a UFG as a function of the complex pairing field $\Delta$. \label{fig:1} }
\end{figure}

The Higgs mode in fermionic superfluids is perhaps one of the most intriguing collective excitation modes of such a system, as its characteristics defy our usual concepts about collective motion. If one were trying to stretch or compress adiabatically any system, the work performed would be interpreted as the potential energy corresponding to that collective degree of freedom. One would have to add to that a corresponding collective inertia and with the emerging collective Hamiltonian one would be able to predict rather accurately the oscillations (sometimes of both small and large amplitude) around the ground state equilibrium configuration. In the case of a fermionic superfluid one characterizes the work performed by an adiabatic ``external'' pairing field $\Delta_{ext}(\vec{r},t)$ with a potential energy corresponding to this collective mode having the shape of a ``Mexican hat," e.g. the energy of a superconductor just below the critical temperature in the Landau-Ginsburg phenomenological model. The ``Mexican hat" potential is ubiquitous in describing various symmetry breaking mechanisms.  The Anderson-Bogoliubov sound waves, with a dispersion $\omega = ck$ \cite{anderson,degennes}, represent the Goldstone modes which appear as a result of the broken gauge symmetry in the case of a superfluid. The Goldstone modes are described by waves of the phase of the pairing field, which propagate along the bottom of the Mexican hat potential and require a very small energy, see Figure (\ref{fig:1}). The oscillations of the phase of the pairing field are accompanied by small oscillations of the number density as well. In the case of a UFG the sound velocity is given by $c =\sqrt{\xi}\hbar k_F/\sqrt{3}m$, which differs from the free Fermi gas only by the factor $\sqrt{\xi}<1$. The Higgs mode on the other hand corresponds to radial oscillations of the pairing field along the radius in the Mexican hat potential.  Naively one would expect that if the pairing field is excited in the radial direction with a small amplitude, one would observe radial oscillations of the pairing field with a finite frequency (unlike sound modes, whose frequency tends to zero in the long wave-length limit), predicted to be exactly $\hbar\omega = 2\Delta_0$ \cite{anderson}, where $\Delta_0$ is the the ground state value of the pairing field. It was realized later on however that this is not correct,  and Volkov and Kogan \cite{volkov} have shown that the oscillations of the pairing field couple with excited quasi-particles with energies above the ``new" gap $2\Delta_\infty < 2\Delta_0$, and that leads for large times to oscillation of the pairing field  $\Delta (t) = \Delta_\infty+A\sin( 2\Delta_\infty t +\phi)/\sqrt{\Delta_\infty t}$. In this case the amplitude decreases very slowly $\propto 1/\sqrt{t}$ evolves towards a new equilibrium state with a smaller pairing gap $\Delta_\infty < \Delta_0$ and a certain fraction of excited quasi-particles.

\begin{figure}
\includegraphics[width=1.1\textwidth]{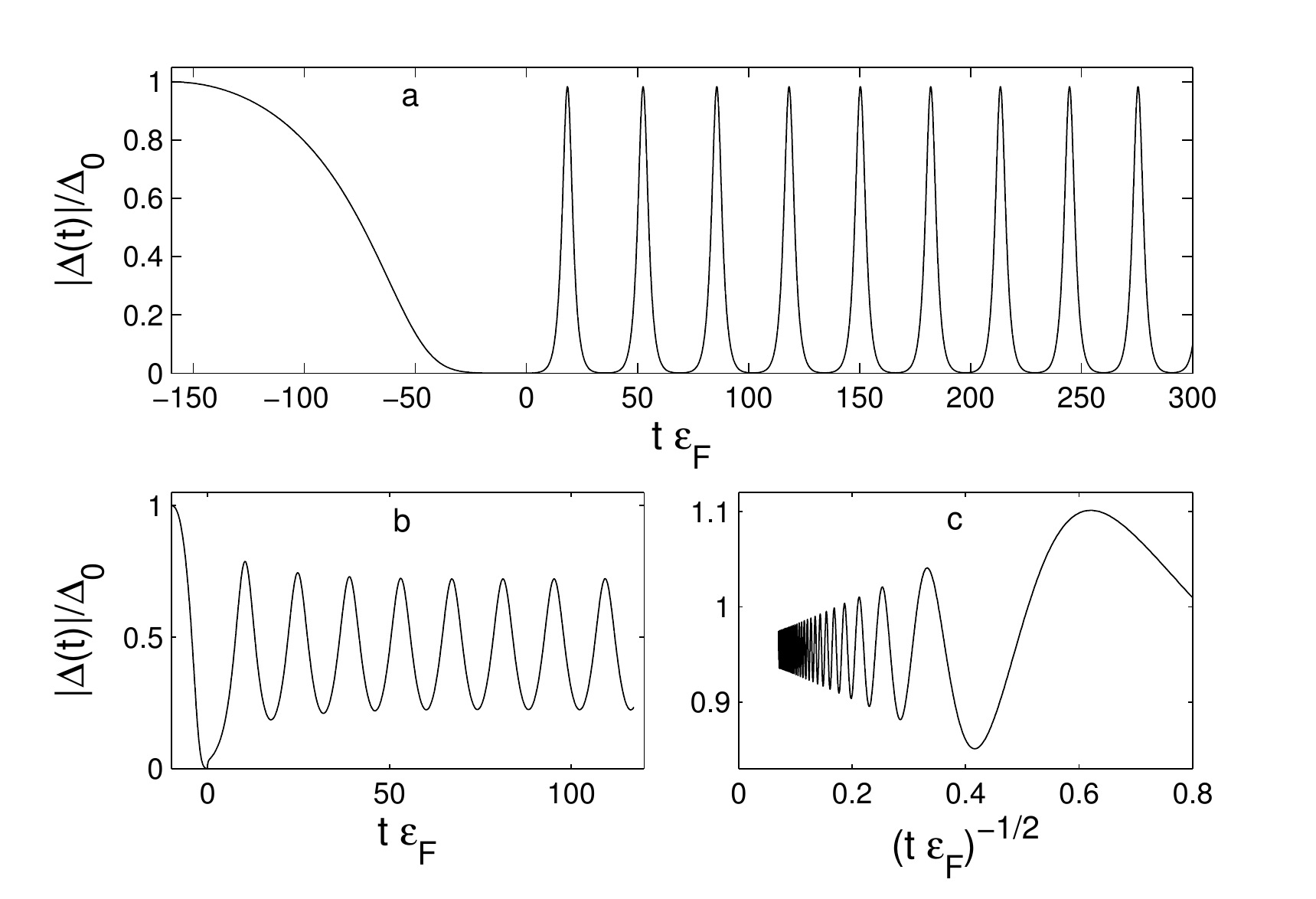}
\caption{The panels $a, b$ and $c$ display response of the homogeneous system to various switching time intervals, see Ref. \cite{higgs} for more details. The initial state is the ground state of the UFG, for times $t<0$ the coupling constant is slowly taken towards the BCS limit in the case shown in panels $a$ and $b$ and at $t=0$ the unitary value of the coupling constant is restored in a very short time interval. \label{fig:2}  }
\end{figure}

\begin{figure}
\includegraphics[width=0.75\textwidth]{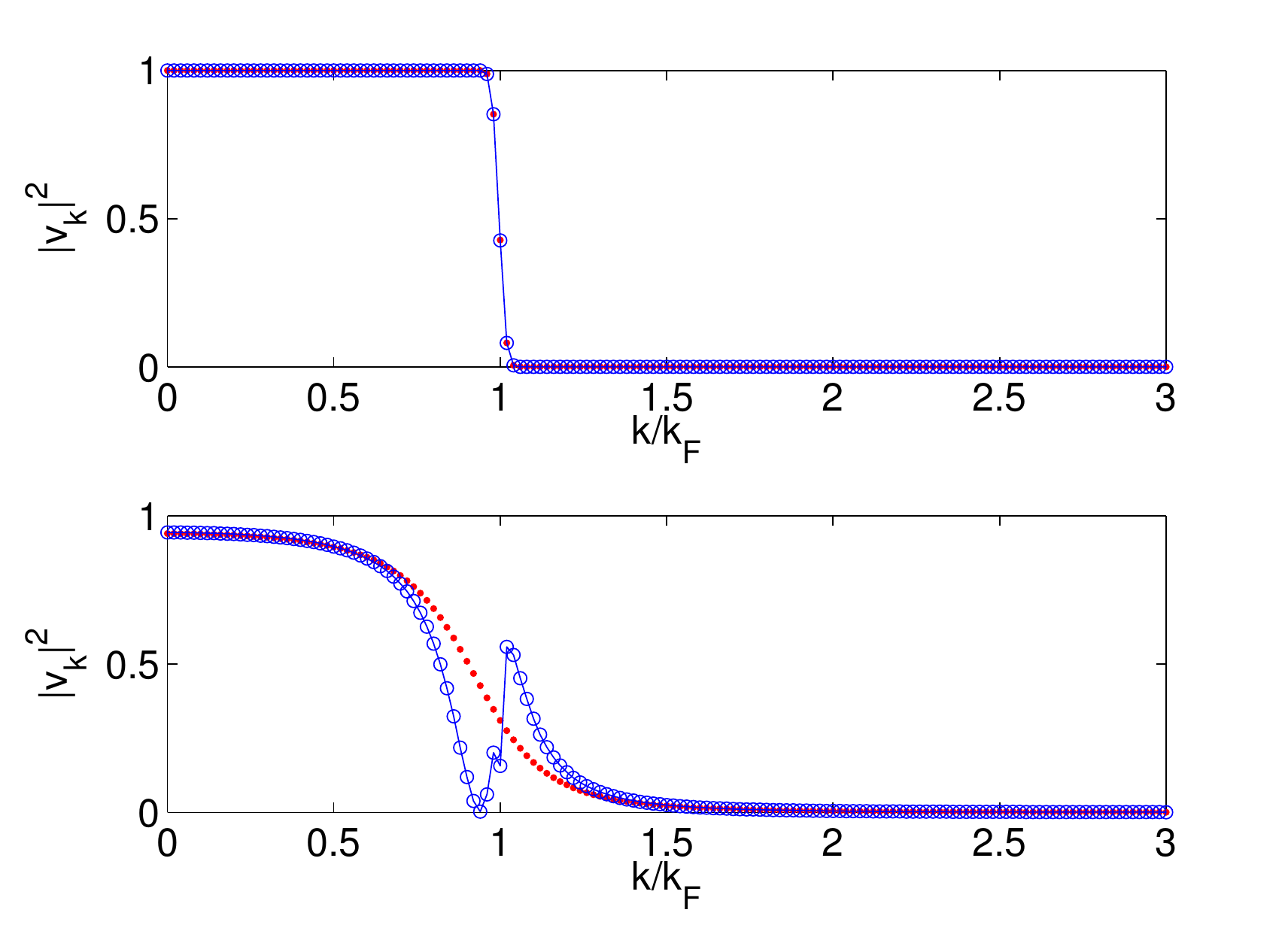}
\includegraphics[width=0.75\textwidth]{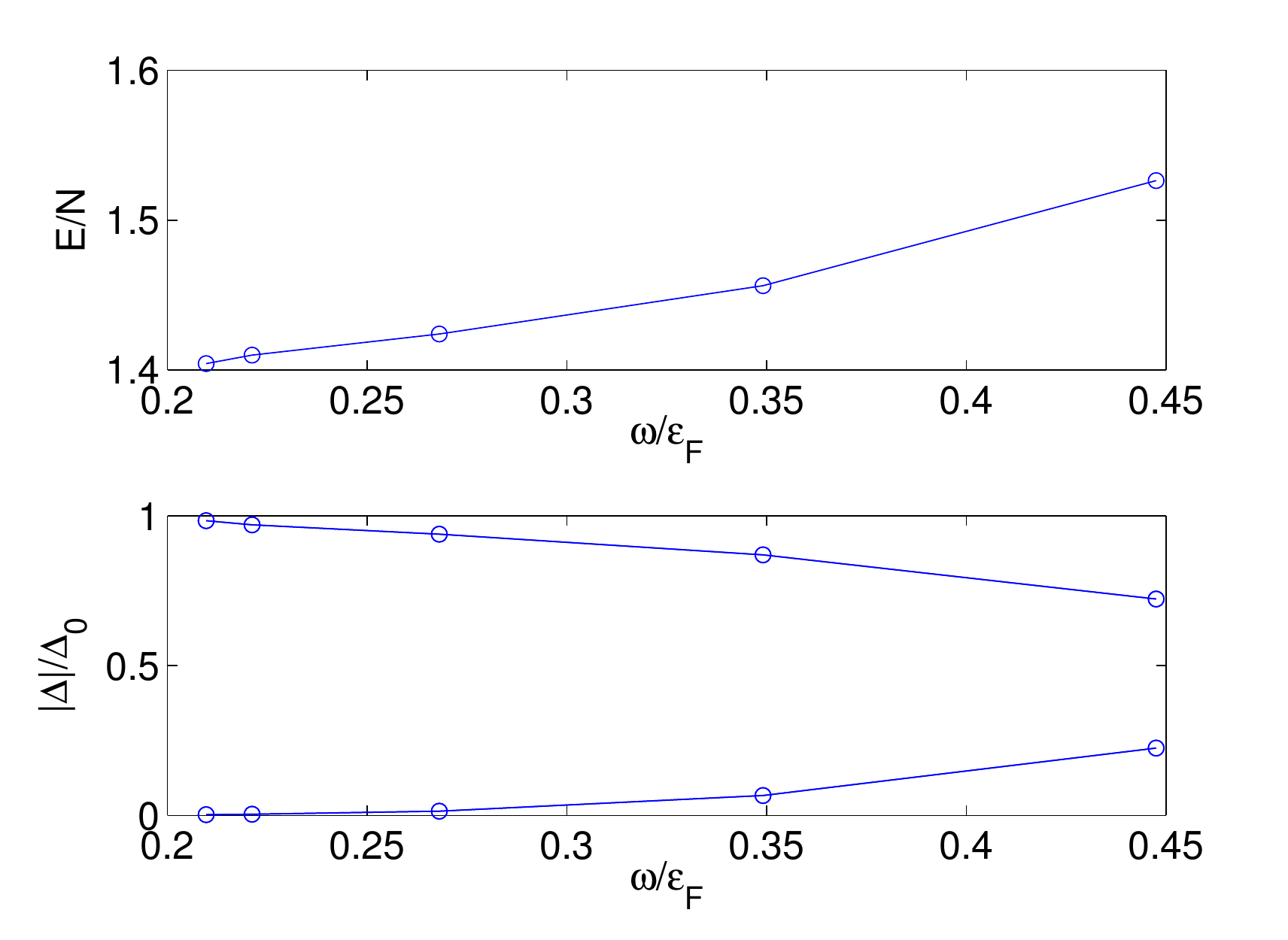}
\caption{ The top two panels display the instantaneous occupation probabilities of the mode shown in upper panel of Figure \ref{fig:2} corresponding to times $t>0$ when the pairing field is at its minimum and maximum values respectively by a solid blue line with circles. With red dots we plotted the equilibrium occupation probabilities corresponding to the same instantaneous values of the pairing gap. In lower two panels we show the maximum and minimum values of the oscillating pairing field and the corresponding excitation energy as a function of the frequency of the Higgs-like modes. \label{fig:3}  }
\end{figure}

These results \cite{volkov}  were obtained for Fermi superfluids in the weak coupling limit, when the oscillations of the pairing field lead to no changes in the quasi-particle self-energy. In a UFG however any change in the pairing field induces a large change of the quasi-particle self-energy, since
\begin{equation} 
h (\vec{ r},t)= -\frac{\alpha \vec{\nabla}^2}{2}+ \frac{\beta(3\pi^2n(\vec{r},t))^{2/3}}{2} -\frac{|\Delta(\vec{r},t)|^2}{3\gamma n^{2/3}(\vec{r},t)}+U(\vec{r},t). 
\end{equation}
In the case of the Higgs mode however the number density stays constant $n(\vec{r},t)\equiv n_0$, where $n_0$ is the ground state value of the number density,  and the oscillations have an infinite wavelength in a perfectly homogeneous system.  Even though the TD-SLDA equations  have a more complex structure, the dynamics of these modes \cite{higgs} is very similar to that predicted for the weak coupling BCS superfluids \cite{volkov}, see panel $c$ of Figure \ref{fig:2} \cite{higgs}. However, if one excites a radial mode with larger amplitude, one observes a qualitatively different behavior, see panels $a$ and $b$ of Figure \ref{fig:2}. Surprisingly, if the system is taken almost adiabatically out of its equilibrium value $\Delta_0$ towards the tip of the Mexican hat potential and then released, the oscillations of the pairing field resembles a  soliton train in time. Notice that most of the time the system illustrated in panel $a$ of Figure \ref{fig:2} is essentially normal, and only for very brief periods of time the pairing gap attains values almost equal to the equilibrium one. The occupation probabilities of the single-particle states are very close to those of a normal system when $\Delta \ll \Delta_0$, but very different from their equilibrium values when the pairing gap attains values close to $\Delta_0$, see upper two panels of Figure \ref{fig:3}. Perhaps the most surprising feature is that the paring field while it oscillates, it never exceeds the ground state value $\Delta_0$ and it never swings on the other side of the minimum of the Mexican hat potential as one would have naively expected.  The modes displayed in panels $a$ and $b$ of Figure  \ref{fig:2} are truly nonlinear, non-equilibrium (see panels $a$ and $b$  Figure \ref{fig:3}), and their frequency depends strongly on the oscillation amplitude, see the panels $c$ and $d$ of Figure \ref{fig:3} \cite{higgs}. These are at the same time very slow modes, with frequencies well below the pairing gap $\hbar \Omega_H < 2\Delta_0$, but at the same time they are truly large amplitude collective modes, not only because of the size of the oscillation amplitude, but also because their excitation energy is equally large, see panel $c$ in Figure \ref{fig:3} \cite{higgs}.  In this respect these modes are somewhat similar to the very large amplitude oceanic waves, which also have very long wavelengths and even though carry a lot of energy, take a long time to dissipate into heat. The results of the somewhat related MIT experiment \cite{magnetic} suggests that the damping of these modes, due to the decay into other modes, is much smaller than one might naively expect at unitarity, related to the fact that the shear viscosity of the UFG is one of the lowest encountered in Nature \cite{cao,viscosity,son,zwerger}. A phenomenological Gross-Pitaevskii like description using  Equation \ref{eq:GP} for example, where $\Psi(\vec{r},t)$ would describe the order parameter $\propto \Delta(\vec{r},t)$, would fail spectacularly. Since in this case there is no coordinate dependence (and in this case $\Delta_{ext}(\vec{r},t)\equiv\Delta_{ext}(t)$), the wave function $\Psi(t)$ would simple have a phase evolution only and its magnitude would never change in time, unlike the behavior in Figure \ref{fig:2} \cite{higgs}. The mechanism by which the Higgs modes dissipate their energy is still an open question, and so far they have no been put in evidence in experiments. A potential decay mechanism was suggested in Reference \cite{dzero}. In inhomogeneous systems the Higgs mode couples to density oscillations \cite{higgs} and that is perhaps the best way to put them in evidence in cold atom  systems. 

\subsection{Nucleation and dynamics of quantized vortices} 


Many more example of numerical experiments  of vortex nucleation and dynamics discussed in this section can be found in Reference \cite{vortices} and especially on the webpage \cite{qmbnt}, where more than 4 hours of videos are available. The UFG is a very interesting system when it comes to its dynamics for one more reason, the Landau critical velocity, the flow velocity at which according to Landau a superfluid looses its superfluidity, is the largest known for any superfluid in appropriate units. There are two mechanisms relevant to our discussion which can lead the the loss of superfluidity in a Fermi superfluid. The first mechanism is due to the excitation of sound modes or phonons. The second mechanism is due to the breaking of Cooper pairs, a mechanism which is absent in Bose superfluids. These two mechanisms have an opposite behavior as a function of the coupling constant, or as a function of the inverse scattering length (as usually is discussed in cold Fermi gases). While one changes the scattering length across the Feshbach resonance \cite{giorgini,bloch,ketterle,grimm,luo,zwerger} from the BCS limit to the BEC limit, the binding energy of a Cooper pair increases monotonically. On the other hand, the sound velocity has an opposite behavior, it decreases monotonically when going form the BCS to the BEC limit. As a result, the maximum critical flow velocity attains its maximum value approximately at unitarity \cite{sensarma,combescot}, namely $v_c \approx 0.37 v_F$, where $v_F = \hbar k_F/m$ is the Fermi velocity of a corresponding free Fermi gas of the same density.  In the following discussion of the vortex dynamics Landau's critical velocity will play an important role. In particular, the fact that Cooper pairs can and do break up during the dynamics of a fermionic superfluid will play also a significant role in our discussion. TDSLDA, unlike traditional approaches of superfluid dynamics, naturally allows for Cooper pairs to break up when conditions are met. 

We will discuss a class of numerical experiments in which the UFG superfluid in its ground state is contained in a vessel with cylindrical symmetry and periodic boundary conditions along the cylinder axis. The type of containment is realized with an almost flat potential and smooth cylindrical walls, high enough to prevent the superfluid from ever spilling over. In the first set of experiments we introduce adiabatically a ``quantum stirrer," parallel to the axis of the cylinder. Such a stirrer can be realized experimentally by a blue detuned laser as in the MIT experiment \cite{zwierlein}. The stirrer rotates with constant angular velocity at a radius close to the edge of the system and we vary its velocity from very slow to supercritical. After stirring the system for a while the stirrer is extracted adiabatically out of the system and the superfluid is left to evolve by itself for a while. While the superfluid is stirred energy is pumped into the system mostly as rotational energy stored in newly formed vortices, and part of this excitation energy as well as in sound waves. This geometry allows us to simplify the numerics by choosing the qpwfs as follows:
\begin{equation}
u_n(\vec{r},t)\Rightarrow\exp(ik_{nz}z)u_n(x,y,t),\quad  v_n(\vec{r},t)\Rightarrow\exp(ik_{nz}z)v_n(x,y,t), \label{eq:uv}
\end{equation} 
where naturally $k_{nz}$ is quantized, due to periodic boundary conditions along the cylinder axis.

\begin{figure}
\includegraphics[width=0.32\textwidth]{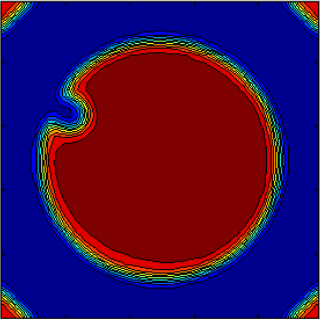}  
\includegraphics[width=0.32\textwidth]{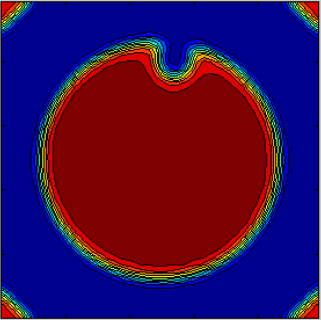}  
\includegraphics[width=0.32\textwidth]{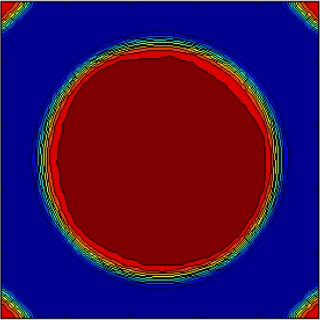} 
\includegraphics[width=0.32\textwidth]{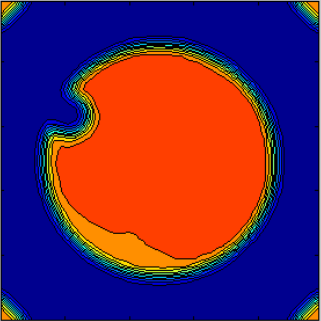} 
\includegraphics[width=0.32\textwidth]{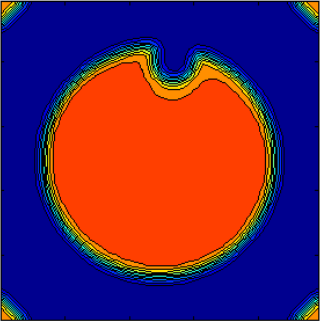}  
\includegraphics[width=0.32\textwidth]{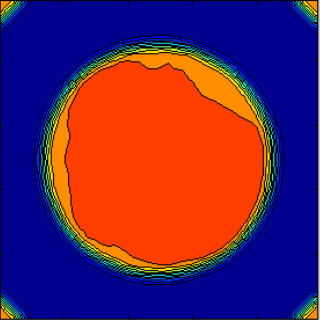}
\caption{The magnitude of the pairing field ($ | \Delta | $, top row) and the corresponding number density ($n$, bottom row) for a UFG system composed of 1800 particles in a $48^{3} $ lattice stirred at $0.45 v_{c}$. No vortices are formed during stirring when the system is stirred at a small velocity, however a few sound modes are excited.\label{fig:4}}
\end{figure}

\begin{figure}
\begin{tabular}{ccc}
\includegraphics[width=0.3\textwidth]{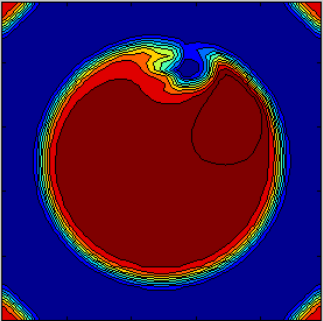} &
\includegraphics[width=0.3\textwidth]{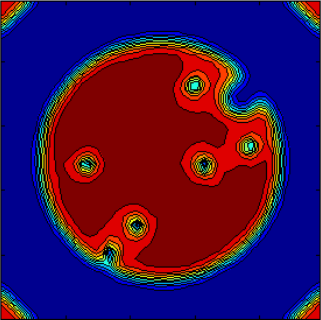} &   
\includegraphics[width=0.3\textwidth]{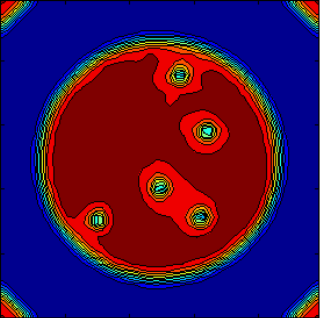}  \\
\includegraphics[width=0.3\textwidth]{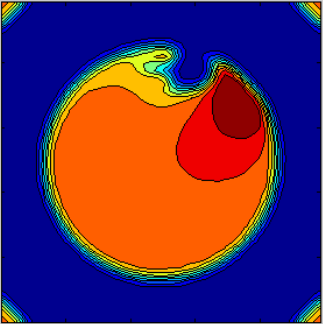} &
\includegraphics[width=0.3\textwidth]{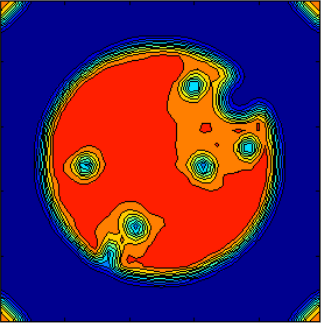} &  
\includegraphics[width=0.3\textwidth]{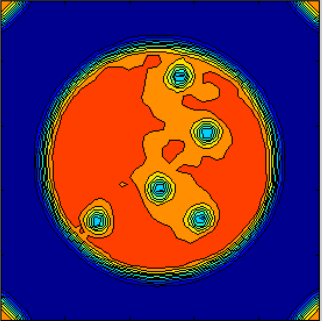}
\end{tabular}
\caption{Several frames showing the UFG at the beginning of the stirring process, during the stirring when several vortices were formed, and at the end, after the stirrer was extracted and the system was left to evolve by itself. The stirring is performed clockwise and the notch close to the boundary of the system shows the instantaneous position of the stirrer. Similar situations are illustrated in Figures \ref{fig:6} and \ref{fig:7}.  The magnitude of the pairing field ($ | \Delta | $, top row) and the corresponding number density ($n$, bottom row) for a UFG system composed of 1800 particles in a $48^{3} $ lattice stirred at a subcritical velocity $0.608 v_{c}$. Five vortices are formed and remain in the system once the stirring concludes. Color coding corresponds to blue, lowest, red intermediate and brown the highest value of the corresponding quantity. \label{fig:5}}
\end{figure}

\begin{figure}
\begin{tabular}{cccc}
\includegraphics[width=0.24\textwidth]{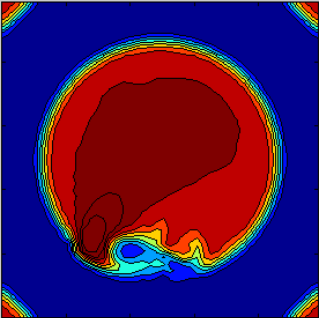}  &
\includegraphics[width=0.24\textwidth]{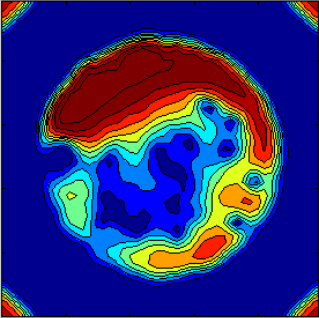}  &
\includegraphics[width=0.24\textwidth]{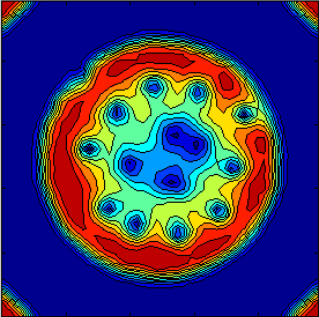}  & 
\includegraphics[width=0.24\textwidth]{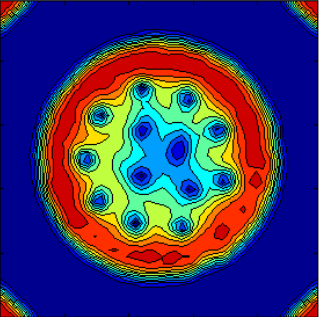} \\
\includegraphics[width=0.24\textwidth]{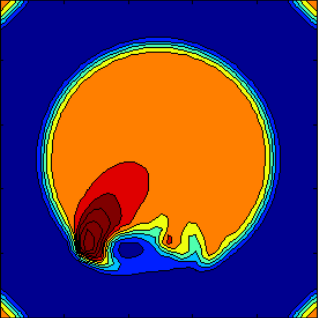} &  
\includegraphics[width=0.24\textwidth]{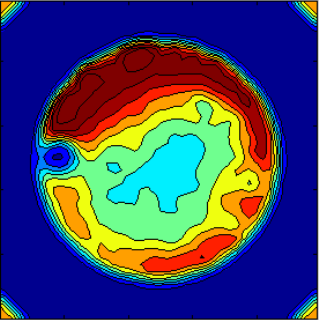}  &
\includegraphics[width=0.24\textwidth]{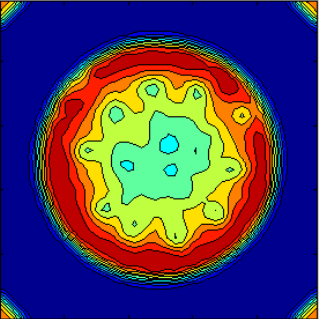}  &
\includegraphics[width=0.24\textwidth]{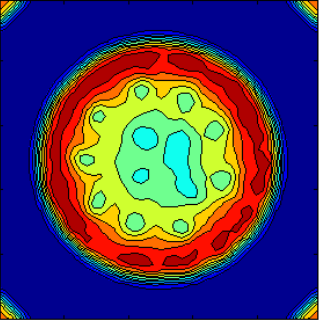} 
\end{tabular}
\caption{The magnitude of the pairing field ($ | \Delta | $, top row) and the corresponding density ($n]$, bottom row) for a UFG system composed of 1800 particles in a $48^{3}$ lattice stirred at supercritical velocity $1.216 v_{c}$. Here thirteen vortices are formed  once the stirring concludes. \label{fig:6}}
\end{figure}

\begin{figure}
\begin{tabular}{ccc}
\includegraphics[width=0.3\textwidth]{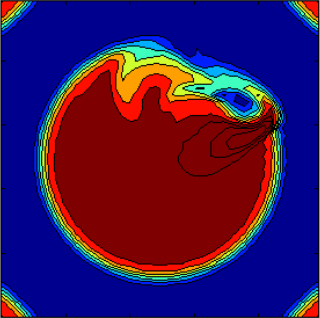}  &
\includegraphics[width=0.3\textwidth]{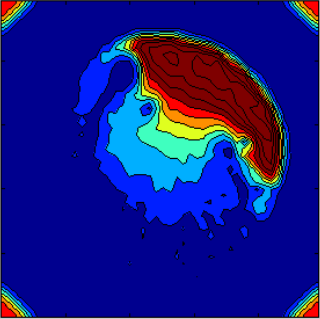}   &
\includegraphics[width=0.3\textwidth]{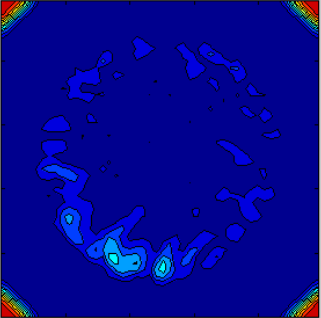}  \\
\includegraphics[width=0.3\textwidth]{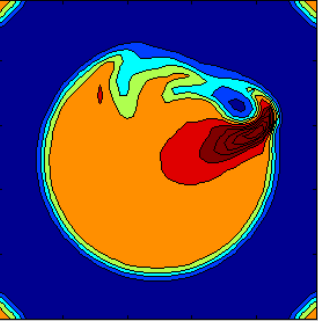} & 
\includegraphics[width=0.3\textwidth]{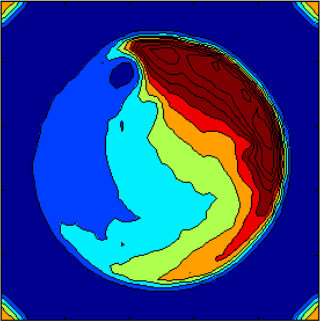}  &
\includegraphics[width=0.3\textwidth]{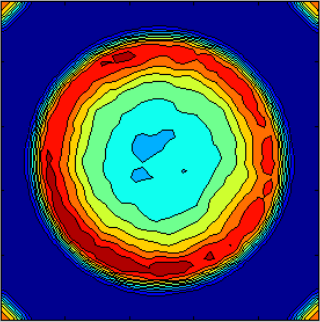} 
\end{tabular}
\caption{The magnitude of the pairing field $ | \Delta |  $ (top row) and the corresponding density $n$  (bottom row) for a UFG system composed of 1800 particles in a $48^{3}$ lattice stirred at supercritical linear velocity $1.824 v_{c}$. No vortices are formed during stirring as the system becomes normal at the end of the run. \label{fig:7}}
\end{figure}

If the stirrer is moving very slowly and not enough energy is pumped into the system so as to create at least one vortex the superfluid typically returns to its initial state at the end of the stirring, see Figure \ref{fig:4}. However, above a certain velocity, which depends on the geometry of the system one or more vortices are created, which tend to form a vortex lattice, constrained by the cylindrical geometry of the container  however, see Figures \ref{fig:5} and \ref{fig:6}.  The energy of a single vortex state depends logarithmically on the radius of the container, and thus is greater in a larger container. Typically sound waves are generated as well, which lead to a non-stationary vortex lattice. Surprisingly, a UFG superfluid can remain superfluid even if stirred at supercritical velocities, as seen in Figure \ref{fig:6}. The explanation is relatively simple, even though the occurrence of this phenomena is somewhat surprising. A UFG superfluid is a gas, and thus highly compressible. During the stirring the gas is gathered by the stirrer as like water by a fast moving paddle, and the local density increases. With the density the local critical velocity increases as well, and the system can remain superfluid even if it is moving with a supercritical velocity of the unperturbed system. When the stirring ends, the system comes to an almost steady state, in which the gas is distributed evenly throughout the container, mostly along the walls. This is not yet a state in which vortices become so close to each other and when a super-vortex state is formed as discussed in Reference \cite{fetter}.  The shape of the container plays an important role in the nucleation and stability of vortices. From the examples shown on the webpage \cite{qmbnt}, one can see that in an ellipsoidal cylinder the vortex lattice is less stable that in a circular cylinder.  In particular we could not generate vortex lattice at as high stirring velocities in an ellipsoidal cylinder as in a circular one.  When the superfluid is stirred at velocities well above the critical velocity, as in Figure \ref{fig:7}, superfluidity is lost, the order parameter monotonically vanishes and the system ends up in a normal state. This is an aspect unique to TDSLDA and absent in any traditional dynamical models of superfluidity. Naturally, when the system becomes normal, the role of dissipative processes is significant and TDSLDA does not describe correctly the dynamics of a normal system, and collisions have to be included, or a totally different theoretical formalism has to be contemplated \cite{long}.

\begin{figure}
\begin{tabular}{cc}
\includegraphics[width=0.3\textwidth,width=2in]{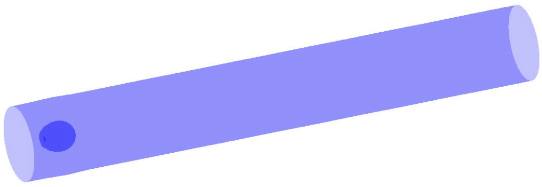} &
\includegraphics[width=0.3\textwidth,width=2in]{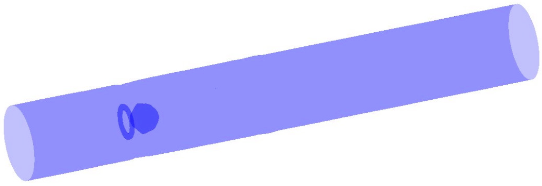} \\
\includegraphics[width=0.3\textwidth,width=2in]{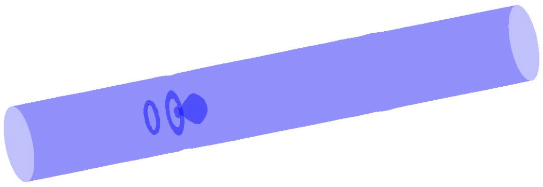}  &
\includegraphics[width=0.3\textwidth,width=2in]{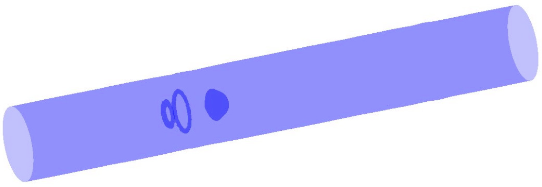} \\
\includegraphics[width=0.3\textwidth,width=2in]{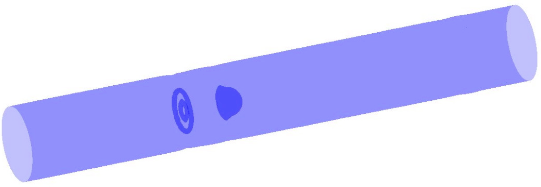}  &
\includegraphics[width=0.3\textwidth,width=2in]{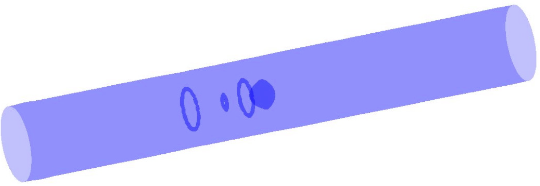}  \\
\includegraphics[width=0.3\textwidth,width=2in]{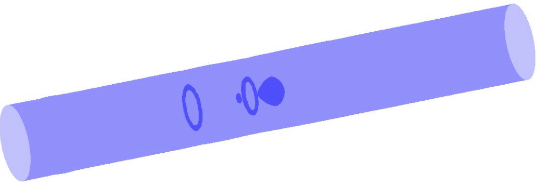} &
\includegraphics[width=0.3\textwidth,width=2in]{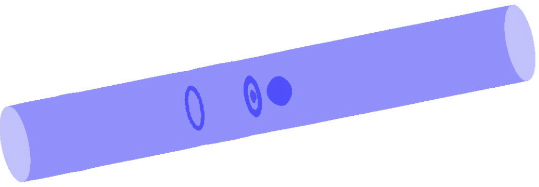} \\
\includegraphics[width=0.3\textwidth,width=2in]{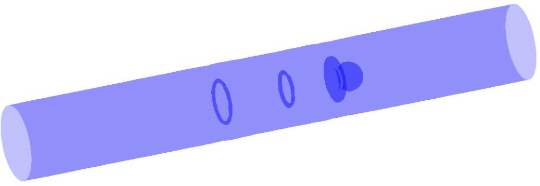} &
\includegraphics[width=0.3\textwidth,width=2in]{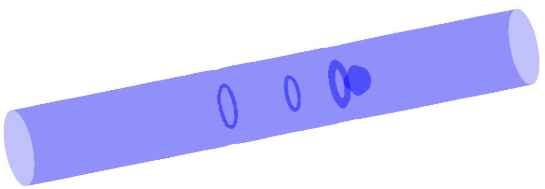} \\
\includegraphics[width=0.3\textwidth,width=2in]{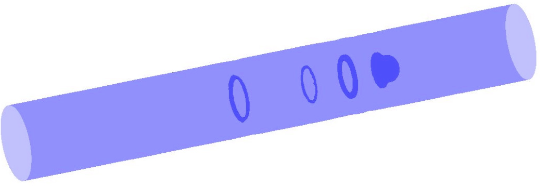} &
\includegraphics[width=0.3\textwidth,width=2in]{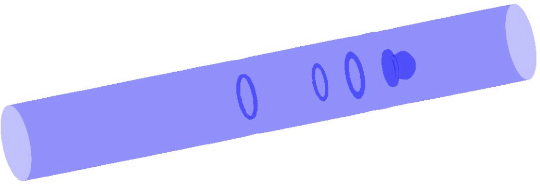} 
\end{tabular}
\caption{The magnitude of the pairing field $|\Delta |$ in UFG system composed on a $32^2 \times 96$ lattice excited with a centered ball directed in the $z$-direction. A sequence of vortex ring formation is observed. In the sequence of figures time progresses left-to-right in a row-major fashion in units of $1/\varepsilon_F$, and similarly in Figure \ref{fig:9}. \label{fig:8}} 
\end{figure}

A genuine 3D numerical experiment, illustrated in Figure \ref{fig:8}, was performed in order to illustrate the generation of vortex rings. An almost impenetrable ball was sent flying along the axis of a very long ``gun barrel" filled with a superfluid UFG at zero temperature.  The velocity of the ball was subcritical $v=0.2v_c$ as well as subsonic. Example of the dynamics generated by a supersonic ball penetrating such a superfluid can be found on the webpage \cite{qmbnt}. In Figure \ref{fig:8} one can see that while traveling the ball is generating perfect vortex rings (mainly due to the symmetry of the problem) of various sizes. As in classical hydrodynamics \cite{lamb,ll}, vortices with larger radii are moving slower than the vortices with smaller radii. An attentive reader will also notice that while ``colliding" two vortex rings do change their radii.  

We have performed a number of numerical experiments to establish microscopically in Fermi superfluids the existence of  the mechanism suggested by Feynman in 1955 \cite{feynman,vinen} to be the origin of quantum turbulence. Turbulence in classical hydrodynamics exists because there always exist strong dissipative processes. Superfluids on the other hand are characterized by negligible viscosity, which in the case of UFG even reaches the minimum conjectured value to exist in Nature \cite{viscosity,cao,son,zwerger}. Feynman conjectured that quantum vortex lines, while moving through the superfluid, can often cross and would likely recombine into new vortex lines. While numerical studies of this mechanism were possible in the case of dilute Bose systems, using the Gross-Pitaevskii Equation \ref{eq:GP} and classical simplified models of vortex lines \cite{schwarz,tsubota} or the two-fluid hydrodynamics \cite{khalatnikov,skrbek}, that was not the case for Fermi superfluids until now. With the exception of the  Gross-Pitaevskii equation, in both the vortex filament models and the two-fluid hydrodynamics vortex quantization is enforced artificially by hand. The first 3D microscopic simulations of the quantum vortex nucleation and dynamics in a fermionic superfluid were performed in Reference \cite{vortices}, where it was demonstrated that 3D quantum vortex lines indeed cross and recombine as Feynman envisioned in 1955 \cite{feynman}.  We illustrate here the incipient mechanism leading to quantum turbulence in a Fermi superfluid with the case of an almost  impenetrable subsonic ball flying parallel to the axis of a very long ``gun barrel" filled with a UFG at zero temperature. While it traverses the tube, the projectile generates a number of imperfect vortex rings  and a number of vortex lines with ends on the edge of the tube. These vortices propagate at various velocities and often they collide with each other and exchange parts of their segments. Sound waves are also excited, either by the projectile or during the vortex crossing and recombination.  In a bigger superfluid volume with a large number of vortex lines these vortex crossings and recombinations lead to quantum turbulence, when the role of viscosity is extremely low.

\begin{figure}
\includegraphics[width=0.32\textwidth]{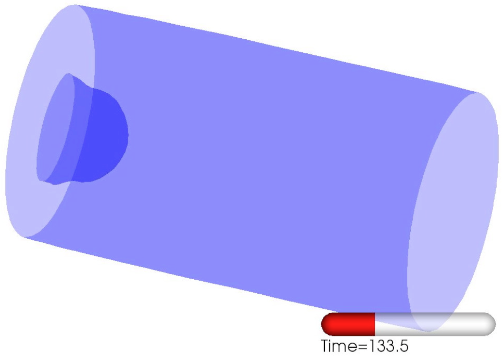} 
\includegraphics[width=0.32\textwidth]{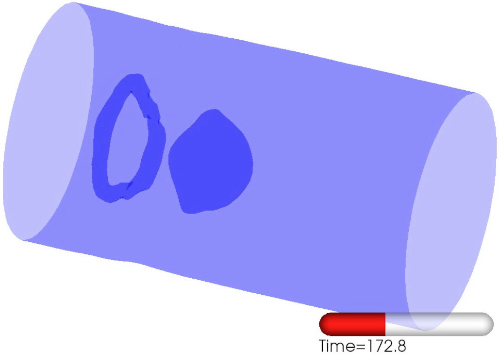} 
\includegraphics[width=0.32\textwidth]{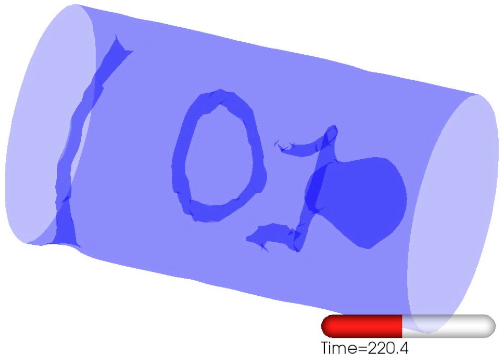} 
\includegraphics[width=0.32\textwidth]{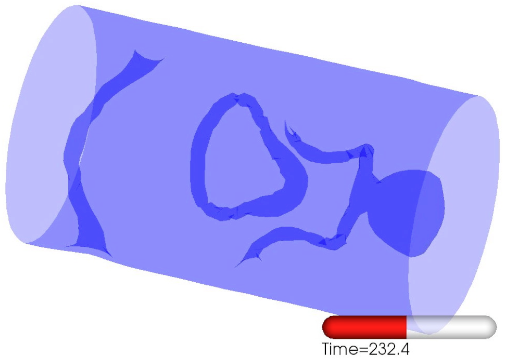} 
\includegraphics[width=0.32\textwidth]{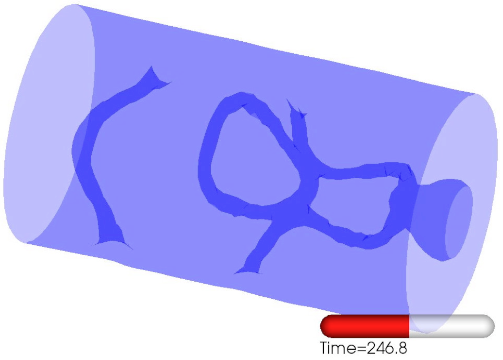} 
\includegraphics[width=0.32\textwidth]{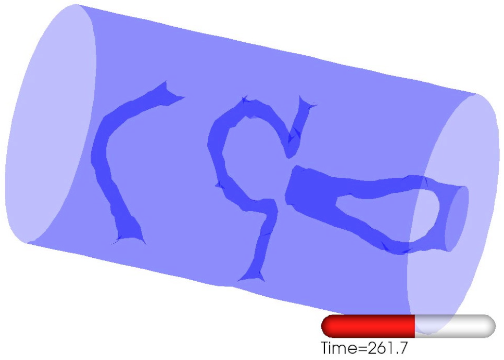} 
\includegraphics[width=0.32\textwidth]{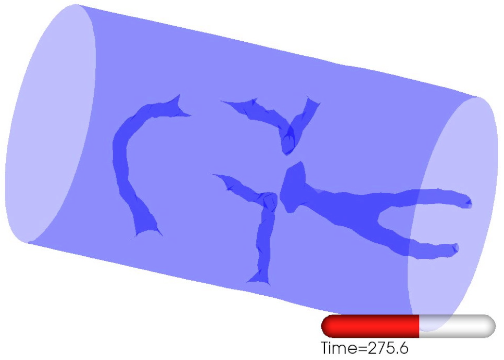}  
\includegraphics[width=0.32\textwidth]{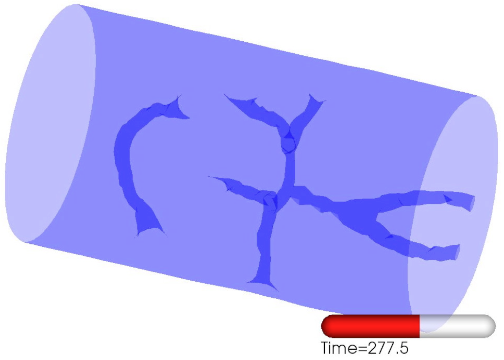} 
\includegraphics[width=0.32\textwidth]{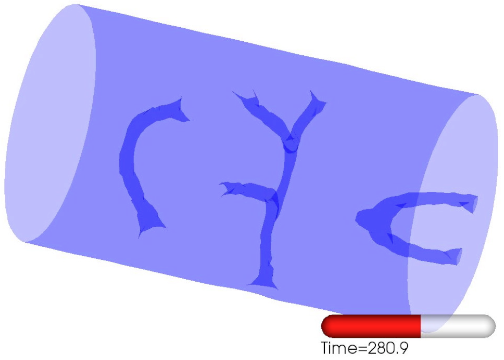} 
\includegraphics[width=0.32\textwidth]{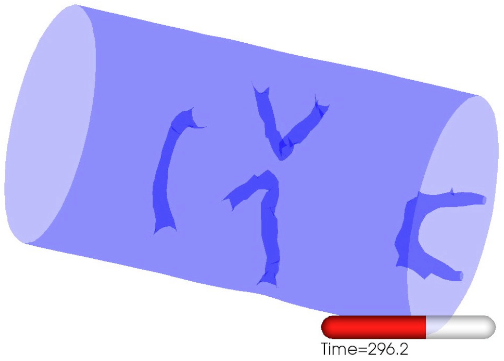} 
\includegraphics[width=0.32\textwidth]{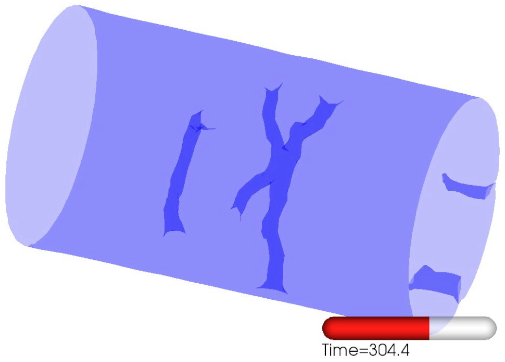} 
\includegraphics[width=0.32\textwidth]{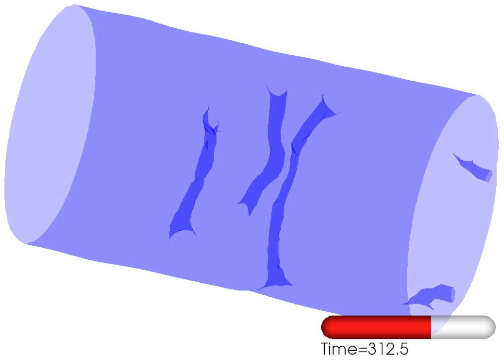} 
\caption{Demonstration of vortex recombination as excited by an off-centered ball flying in the $z$ direction. The crossing and recombination of vortex lines were conjectured to be the origin of quantum turbulence of superfluids by Feynman in 1955 \cite{feynman}. \label{fig:9}}
\end{figure}

\subsection{Quantum shock waves and domain walls}

Shock waves and soliton waves have been studied for more than a century in classical fluid dynamics and plasma physics, and their dynamics and propagation properties depend on the presence of dissipative effects and the subtle interplay of non-linearites and dispersive effects \cite{whitham}.   In the case of UFG moreover, it is already an established fact, both theoretically and experimentally, that shear viscosity is very low, reaching its minimum quantum limiting value \cite{viscosity,cao,son,zwerger}. During the last decade, various experiments performed with dilute cold Bose gases and their theoretical interpretation using the Gross-Pitaevskii Equation \ref{eq:GP} demonstrated that quantum shock waves could be excited in such systems and the role of dissipation is negligible \cite{dutton,simula,hoefer,chang,carretero}. In the recent experiment performed by Joseph {\it et al.} \cite{joseph} with a cloud of cold $^6Li$ fermionic atoms the existence of quantum shock waves in a superfluid Fermi system was confirmed. These authors created a long elongated atomic cloud in a harmonic trap. The cloud was adiabatically cut in two pieces with a laser beam, which was subsequently removed and the two separated clouds accelerated towards each other along the long axis of the harmonic atomic trap. Pictures of the collision revealed that after merging, two shock wave fronts were formed while the cloud was expanding.   The experiment was modeled in Reference \cite{joseph} within a modified hydrodynamic approach, by adding a phenomenological shear viscosity term
\begin{eqnarray}
&&\dot{n } + \vec{\nabla}\cdot (\vec{v} n )=0,  \\
&& n \dot v_k +n\nabla_k\left \{ \frac{|\vec{v}|^2 }{2}  +\mu[n ]+U \right \} +
 \nu \nabla_l \{n   [\nabla \odot v  ]_{kl}\} =0,
\end{eqnarray}
where we have suppressed the arguments $(\vec{r},t)$. One can choose the mass of the atom $m=1$, $v_k$ and $\nabla_k$ are the cartezian coordinates of $\vec{v}$ and $\vec{ \nabla}$ respectively, and $[\nabla \odot v]_{kl} = \nabla_k v_l+\nabla_l v_l-2\delta_{kl}{\mathrm {div}}\,\vec{v}/3$.  Above  $\mu[n]$ is the chemical potential in homogeneous matter at a given number density $n$, and we have suppressed the explicit dependence of the number density $n(\vec{r},t)$, velocity $\vec{v}(\vec{r},t)$ and external trapping potential $U(\vec{r},t)$ on space-time coordinates.  Unlike the case of dilute Bose gases where the quantum shock waves were interpreted as dispersive shock waves \cite{dutton,simula,hoefer,chang,carretero} with no need for dissipative effects, the results of the experiment \cite{joseph} on UFG received an interpretation similar to classical shock waves, when dissipation plays a crucial role in the formation of the shock wave front. These two distinct interpretations of the experiments on Bose and Fermi dilute gases are difficult to reconcile, especially in the light of what has been established so far studying the shear viscosity of these systems \cite{viscosity,cao,son,zwerger}. In the BEC regime, the role of dissipation is negligible and the shock wave and the density ripples identified with soliton trains can be described by dispersive effects alone. Viscosity was introduced phenomenologically in Ref. \cite{joseph} to avert the onset of a ``gradient catastrophe''  \cite{bettelheim}. At the same time one would expect that in an UFG the role of viscosity is even less important than in a BEC system, where viscosity was not needed to model experiments, as the UFG is widely accepted as a prime example of an almost perfect fluid. A significant limitation of the hydrodynamic approach \cite{joseph} is the inability to describe quantum topological excitations (quantized vortices and domain walls in particular), both of which have been observed in the similar experiments with bosons \cite{dutton,simula,hoefer,chang,carretero}.

In the case of colliding UFG clouds, we observe the generation of both quantum shock waves and domain walls, the excitation of which have been suggested for some time in different kind of simulations \cite{antezza,scott,spuntarelli,liao}. The domain walls are excitations of the superfluid order parameter and not the number density ripples identified as soliton trains trailing the wake of the shock waves, as discussed in Refs. \cite{dutton,simula,hoefer,chang,carretero}. We will make this distinction in order to avoid confusion. We show that the number density of two colliding UFG clouds shows a behavior very similar to the one observed in experiment  \cite{joseph}.  In the wake of the quantum shock waves we observe the formation of domain walls.  The domain walls emerge as quite sharp changes in the phase of the superfluid order parameter by $\pi$, and are correlated with minima of the number density. One can distinguish two types of domain walls, Fulde-Ferrell-like \cite{FF} in which the phase of the order parameter changes continuously by $\pi$, and Larkin-Ovchinnikov-like \cite{LO} when the order parameter merely changes signs. Domain walls, which are Fulde-Ferrell-like in this case,  propagate through the system at slower speeds than the quantum shock waves and are topological excitations  similar to quantum vortices. Domain walls always appear in pairs with opposite jumps of the order parameter phase and appear to collide essentially elastically with one another and with the system boundary. These phenomena are observed in the absence of any dissipation, which is expected to play a negligible role at temperatures close to absolute zero and especially in a UFG \cite{viscosity,cao,son,zwerger}.

\begin{figure}
\includegraphics[width=0.48\textwidth]{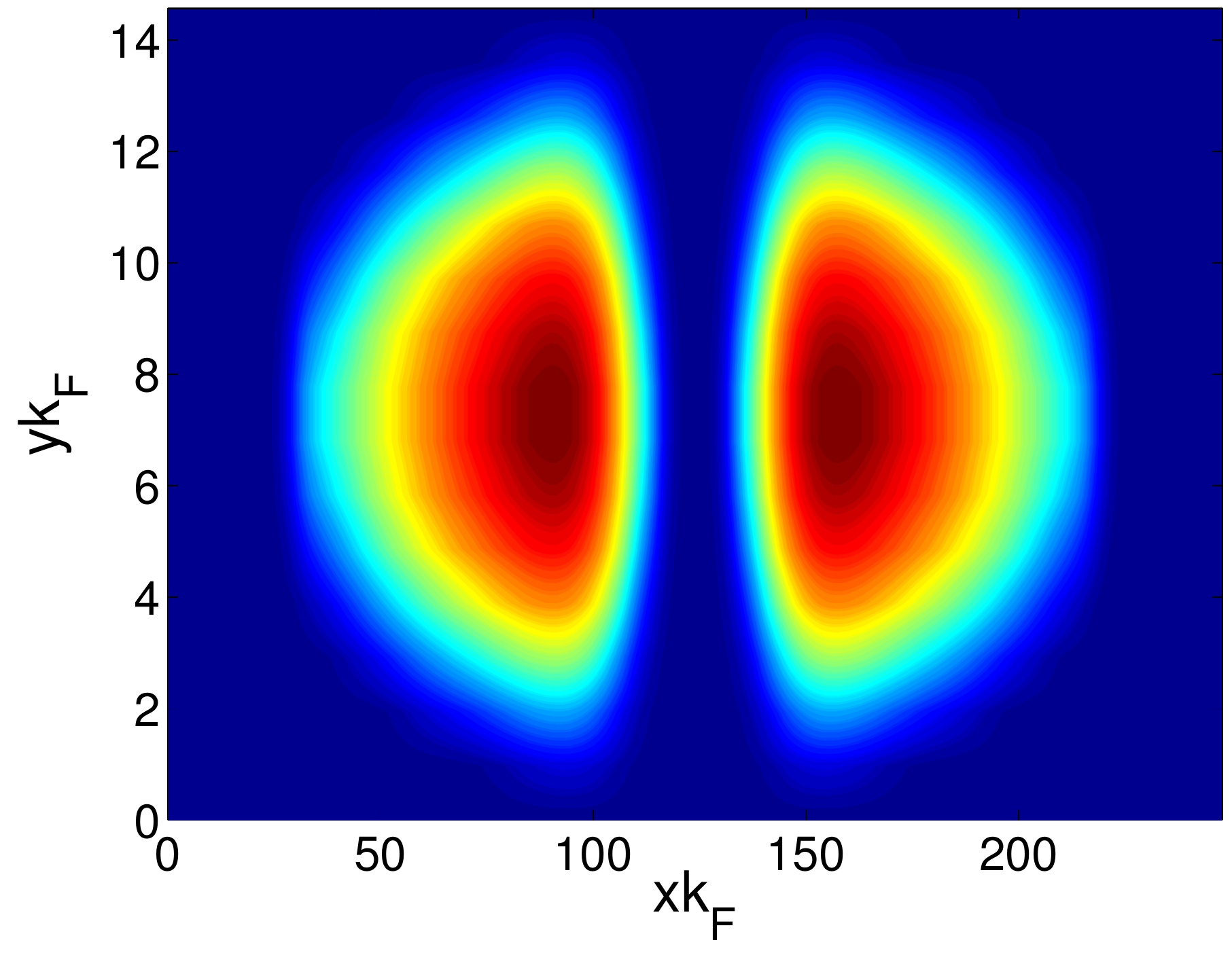}
\includegraphics[width=0.48\textwidth]{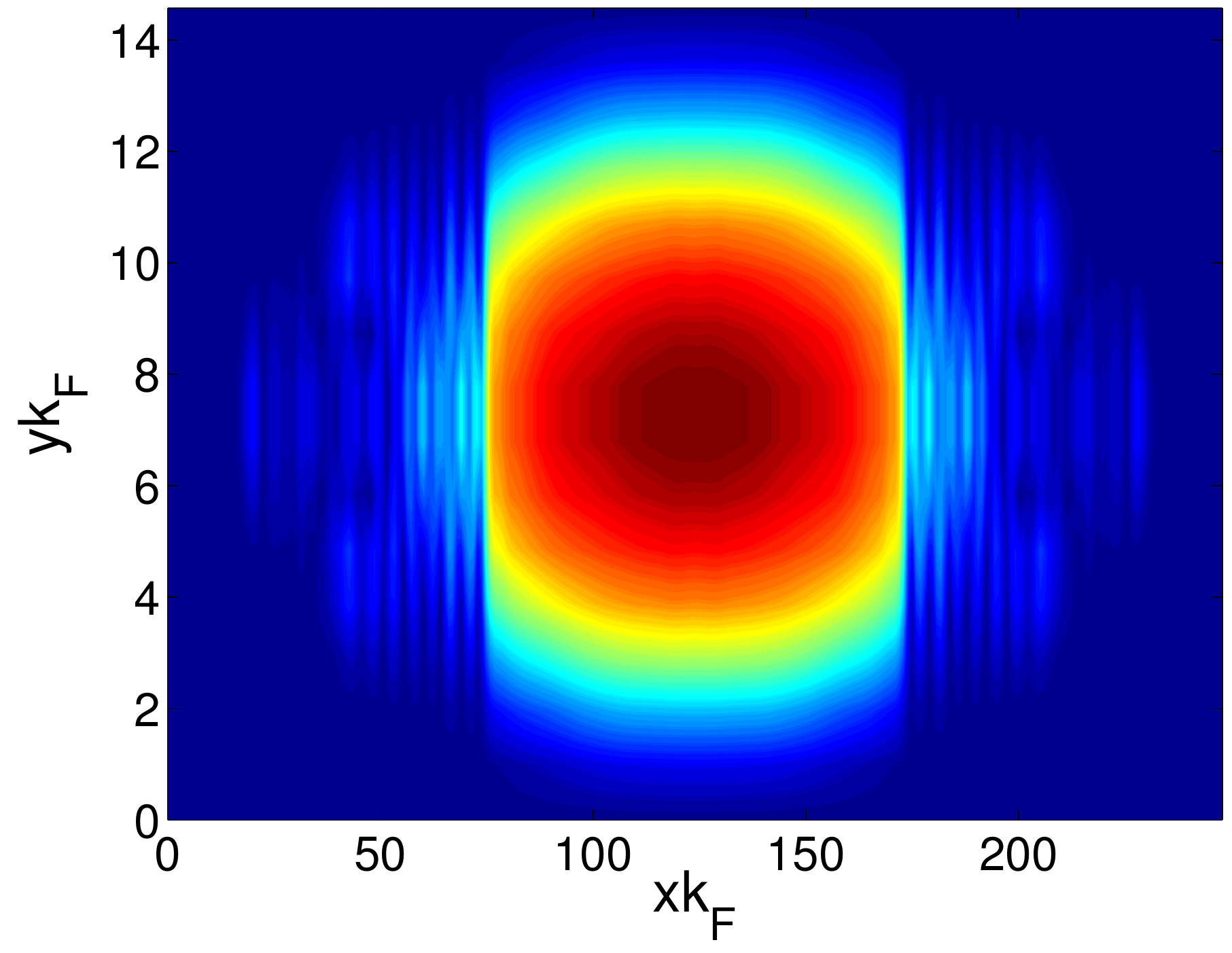}
\caption{ The left panel displays the two cooling clouds  before the collision, and the right panel the expanding cloud formed after the collision respectively. The expansion front is almost flat, similarly to the one observed in the Duke experiment \cite{joseph} and interpreted as a quantum shock wave.  Notice that the visual aspect ratio of the clouds is not shown to scale for display purposes.   \label{fig:10}}
\end{figure}

\begin{figure}
\includegraphics[width=0.32\textwidth]{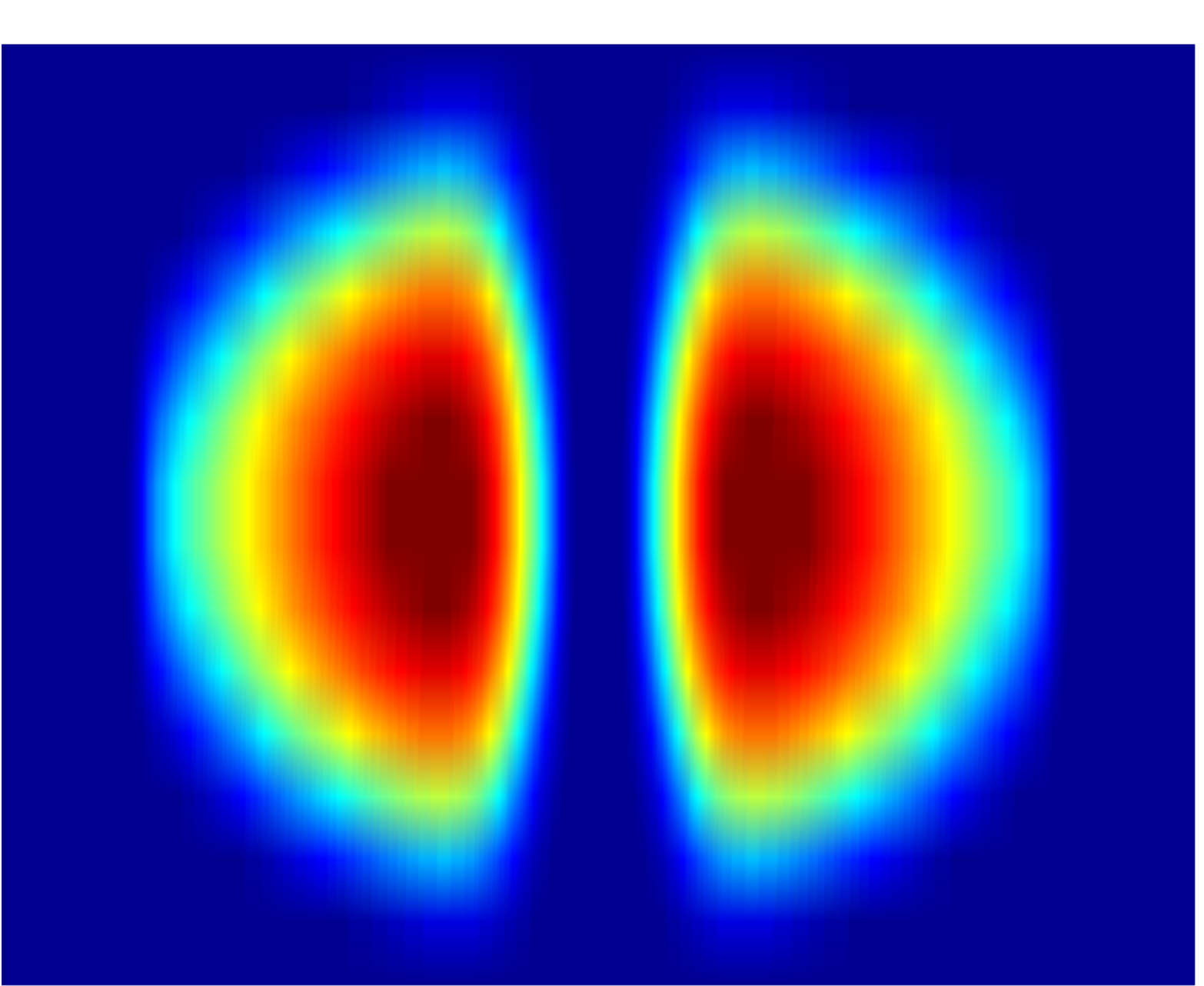}
\includegraphics[width=0.32\textwidth]{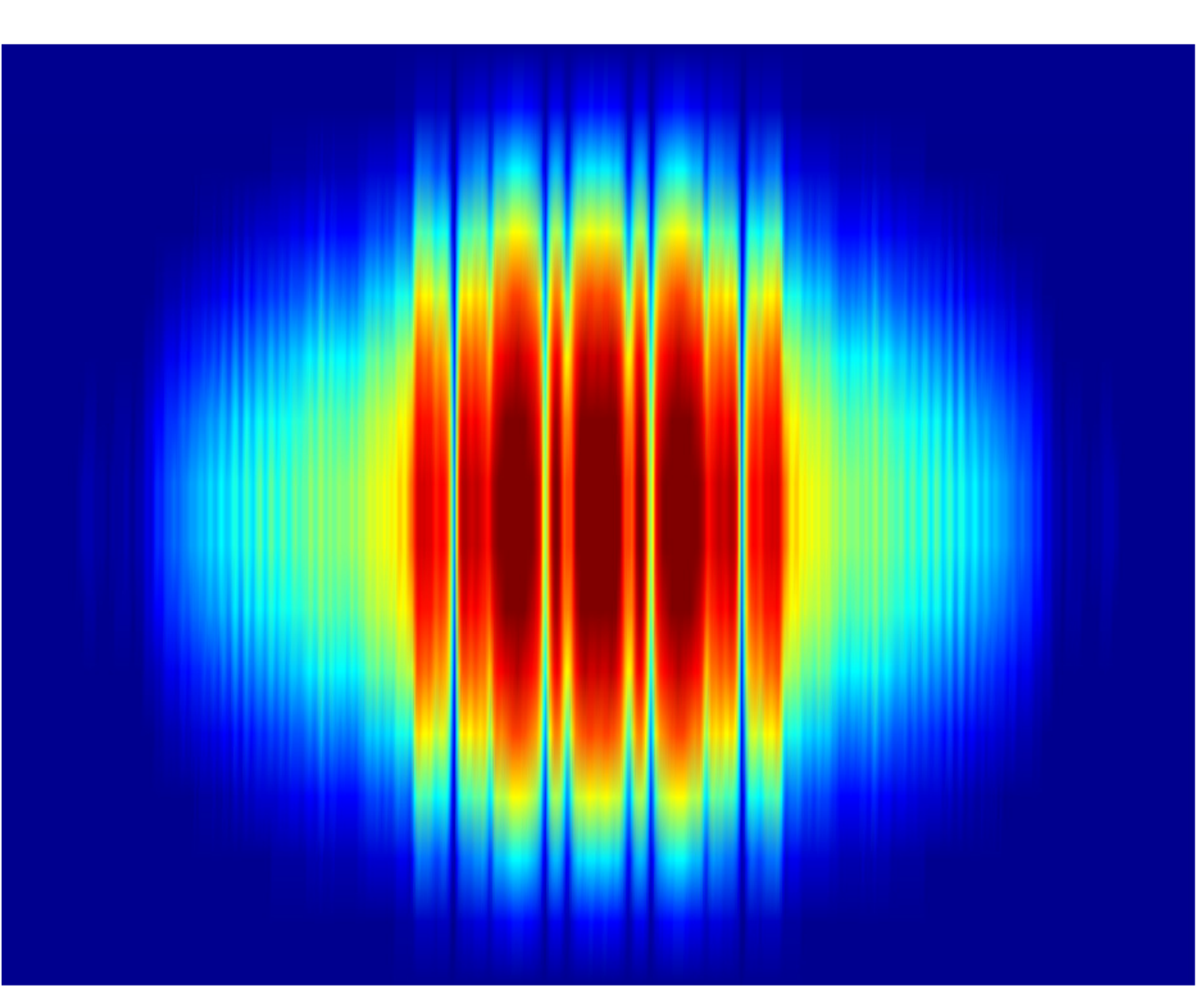}
\includegraphics[width=0.32\textwidth]{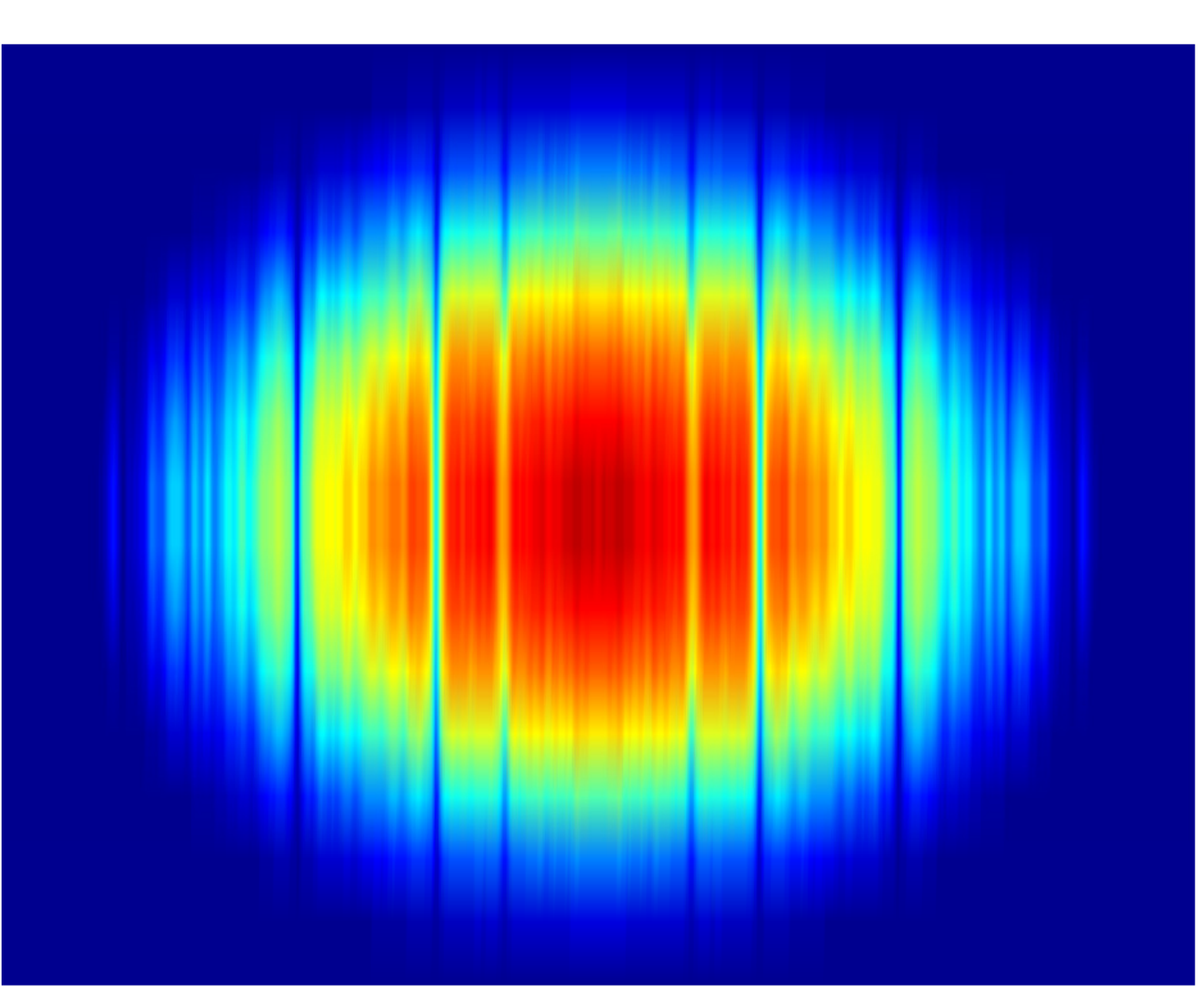}
\caption{ Three consecutive frames showing the absolute magnitude of the pairing field $|\Delta(x,y,t)|$  in the $xy$-plane at times $t\varepsilon_F/\hbar =30, \; 350,\; 690$. The $x$- and $y$-directions (not shown to scale here) have an aspect ratio of $\approx 30$. We have observed the formation of domain walls so far only in traps with elongations larger than in experiment \cite{joseph}.}
\label{fig:11}  
\end{figure}

\begin{figure}
\includegraphics[width=0.81\textwidth]{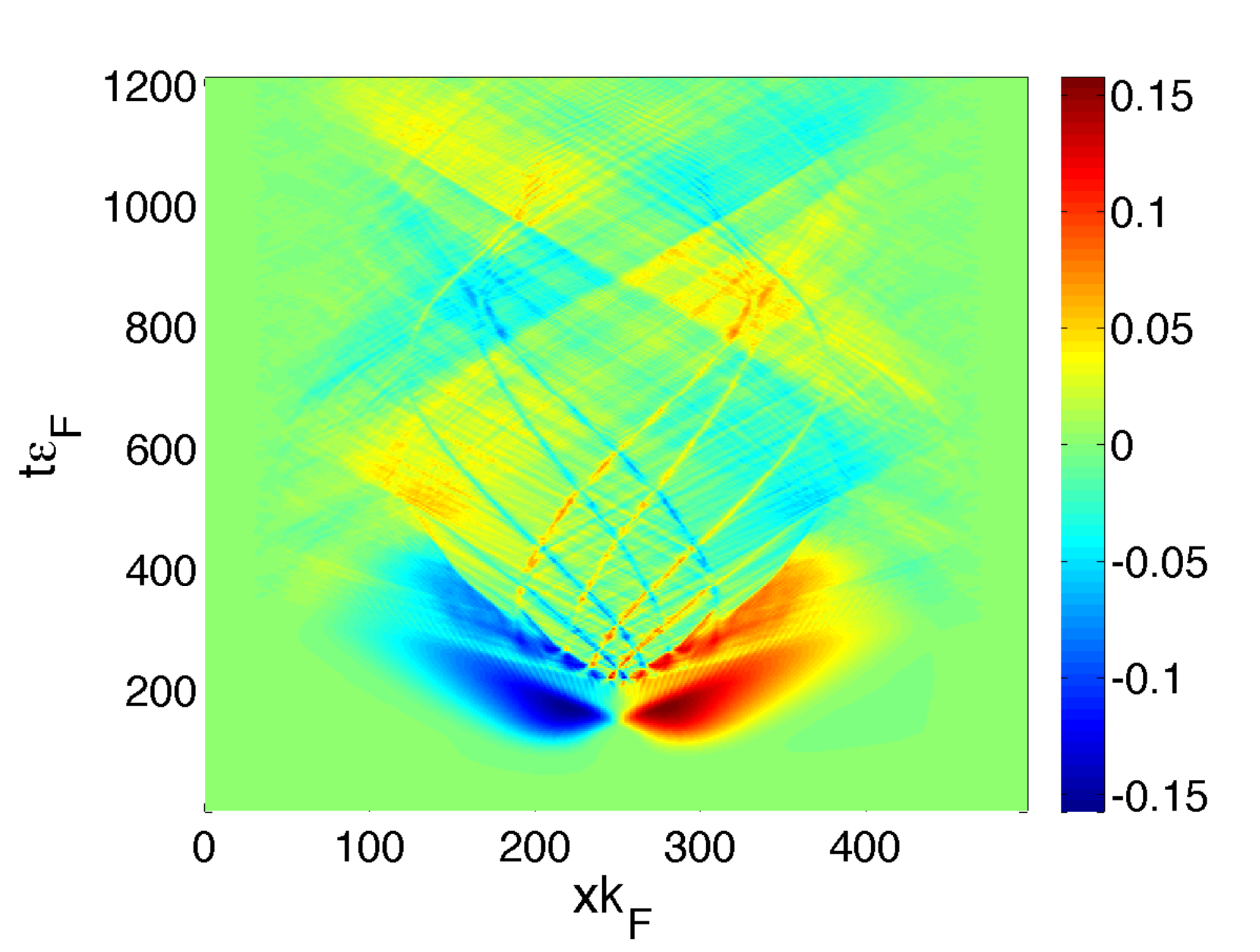}
\caption{ The $x$-component of the collective flow velocity field $v_x(x,0,t)$ along the axis of collision in a space-time diagram. At the front of the two shock waves the velocity field  undergoes a rapid change in sign, and the matter flows in two opposite directions, fully consistent with a shock wave interpretation of the expansion front.   \label{fig:12} }
\end{figure}

\begin{figure}
\includegraphics[width=0.5\textwidth]{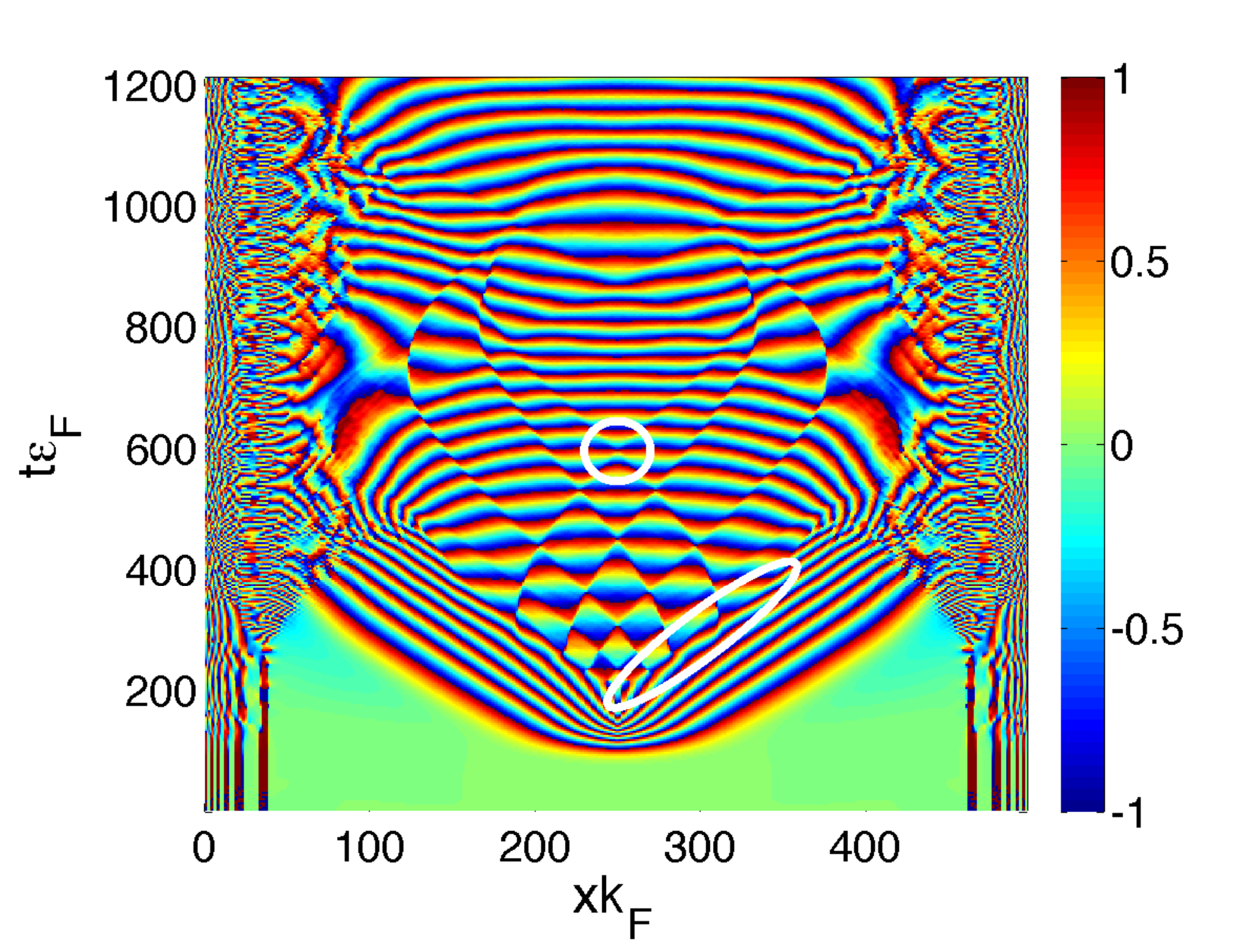}
\includegraphics[width=0.5\textwidth]{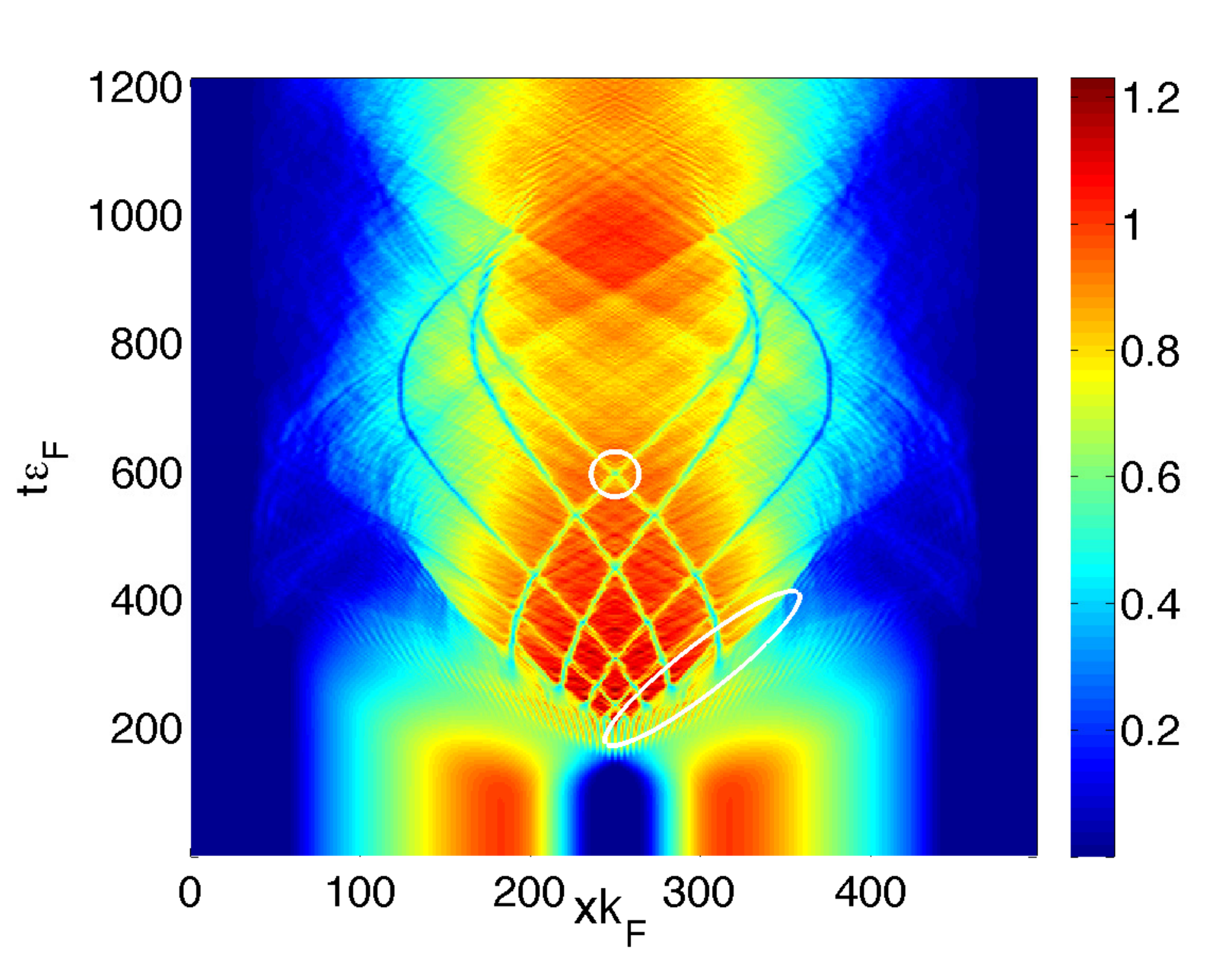}
\caption{ The left panel shows the phase of the pairing field in units of $\pi$. One can clearly see lines of sharp phase change by approximately $2\pi$, consistent with the formation of domain walls. The contour of the density profiles is shown in the right panel. The white circle shows one area where two domain walls experience what appears to be an elastic collision. \label{fig:13} }
\end{figure}

We have performed simulations of the cold atom cloud collisions assuming that qpwfs have the structure  as in Equation \ref{eq:uv},  with periodic boundary conditions in the $z$-direction and a rather stiff  harmonic confining potential in the $y$-direction near the box boundary.  The time-dependent trapping potential along the collision $x$-axis had a similar profile as the one used in experiment \cite{joseph}, namely a shallow confining potential in the $x$-direction with a high potential barrier in the middle that was rather rapidly lowered. The solitons and the shock waves now are two-dimensional in character, their stability properties are slightly different than in 3D \cite{capuzzi}. Typical results of these simulations are shown in Figures \ref{fig:10}, \ref{fig:11}, \ref{fig:12} and \ref{fig:13}.  

The simulation results, Fig. \ref{fig:10}, show remarkable similarities with the experiment \cite{joseph}. The shock wave speed in experiment and simulations agree within $\approx 25-30\%$. The differences can be ascribed to various experimental uncertainties (in particular the particle number and the value of $k_F$ after expansion) and different geometries. In spite of being confined in the $y$-direction in a harmonic potential, the domain walls are planes perpendicular to the collision $x$-axis. However, in Reference \cite{joseph} the experimental set-up prevented the authors from observing the domain walls. The images corresponding to various frames reported there were taken in different realizations of the two colliding clouds. The phase differences of the two initially separated condensates are random and cannot be controlled from one shot to another, similar to collisions between Bose dilute clouds \cite{andrews}. The density profile fluctuations from shot-to-shot  in Reference \cite{joseph} point to a rather low spatial resolution attained in these measurements (see Fig. 2 in Reference \cite{joseph}), which is insufficient to put in evidence domain walls.  We have performed simulations by varying the initial relative phase of the condensates. While the overall picture of the collisions remains unchanged, the number of domain walls created varies. The density ripples in the wake of the shock waves discussed in experiments with Bose dilute clouds  \cite{dutton,simula,hoefer,chang,carretero} and interpreted there as a soliton train, are formed here as well. By zooming in the online Figures \ref{fig:11}, \ref{fig:12} and  \ref{fig:13} one can notice that before the shock wave is formed well defined matter wave interference occurs. The discontinuity in the number density and order parameter at the wake of the shock wave is accompanied by a similar discontinuity in the collective flow velocity, see Figures \ref{fig:12} and \ref{fig:13}.  The domain walls which form in the wake of the shock wave have lower speeds. With a white circle we marked the region where two domain walls collide, apparently elastically. The propagation speed of the domain walls is lower than sound speed.  One can see the scars left by the domain walls also  in the collective flow shown in the upper panel of Figure \ref{fig:12}. In a hydrodynamic approach, with a phenomenological gradient correction term to the EDF, Salasnich and collaborators \cite{salasnich} were able to reproduce the shock waves observed in the Duke experiment \cite{joseph} in the absence of viscosity.  These authors suggest an hydrodynamic approach based on the Lagrangean
\begin{eqnarray}
\mathcal{L}&=& i\hbar \Psi^*(\vec{r},t)\partial_t\Psi(\vec{r},t) -\frac{\hbar^2 |\vec{\nabla}\Psi(\vec{r},t)|^2}{4m} \nonumber \\
&-& \xi \frac{\hbar^2 3(3\pi^2)^{2/3}n^{5/3}(\vec{r},t)}{10m} -U(\vec{r},t) n(\vec{r},t),
\end{eqnarray}
in which they made the identification $\Psi (\vec{r},t) = \sqrt{n(\vec{r},t) /2}\exp[i\chi(\vec{r},t)]\propto \Delta(\vec{r},t) $ and $2m\vec{v}(\vec{r},t) =\hbar \vec{\nabla}\chi(\vec{r},t)$. Rewritten in terms of $n(\vec{r},t)$ and $\vec{v}(\vec{r},t)$ the emerging equations look formally like the hydrodynamic equations at zero temperature, plus the ``quantum pressure" term, which was neglected in Reference \cite{joseph}. In the BEC regime this identification amounts to using the Gross-Pitaevskii Equation \ref{eq:GP} for the dimers/fermion pairs, after replacing the interaction energy as well with the dimer-dimer repulsion. In the BCS limit this identification is clearly wrong, as well as at unitarity. The hope is that qualitatively one can encapsulate the dynamics reasonably well however. These authors also observed the formation of density ripples at the front of the shock wave, similar to those observed in our simulations, and also in the case of quantum shock waves in Bose systems \cite{dutton,simula,hoefer,chang,carretero}. An analytic explanation of the origin of these density ripples in 1D Fermi systems was given recently by Bettelheim and Glazman \cite{eldad}. 

\subsection{Nuclear collective modes} 

The isovector giant dipole resonance (GDR) is perhaps the simplest example of a nuclear collective motion of all the protons against all the neutrons. Since its observation in the photo-absorption cross section \cite{bothe}, it has been intensively studied as it combines several challenging aspects of the physics of the atomic nucleus \cite{levinger,harakeh}. Even though the GDR is practically harmonic in character,  it is not an adiabatic collective mode and various damping mechanisms of the collective energy are at work  \cite{bertsch}. In the early models of Migdal \cite{migdal}, Goldhaber-Teller \cite{GT}, and Steinwedel-Jensen \cite{SJ} the GDR is described as the relative motion of two fluids, either compressible or incompressible,  with neutrons and protons vibrating around a common center of mass, and the mass dependence of the excitation energy reads $A^{-1/6}$ and $A^{-1/3}$ respectively \cite{ring}.  A good estimation of the GDR vibrational frequency is $\hbar \omega \approx 80$ MeV $A^{-1/3}$ for spherical nuclei. The GDR is interpreted simply as the equivalent of the zero-sound in a nuclear system and the size of the nucleus sets a constraint on the largest wavelength. In the case of  deformed nuclei, the GDR peak is split, with various frequencies revealing different principal axes of the nuclear shape. The total width of the GDR is mainly due to a couple of mechanisms:  the coupling of the GDR to complex nuclear configurations $\Gamma^{\downarrow}$, and the coupling to the continuum, leading to the escape of neutrons and protons  $\Gamma^{\uparrow}$. These two widths contribute to the total width of the GDR, $\Gamma = \Gamma^{\downarrow} + \Gamma^{\uparrow}$, and their relative contributions vary depending on the mass number $A$ and the $N/Z$ ratio.  The escape width is typically more important for light nuclei. The physical mechanisms related to $\Gamma^{\downarrow}$ may be quite complicated  and involve coupling to low energy surface vibrations, Landau damping and collisional damping \cite{bbb}.

\begin{figure}
\includegraphics[width=0.75\textwidth]{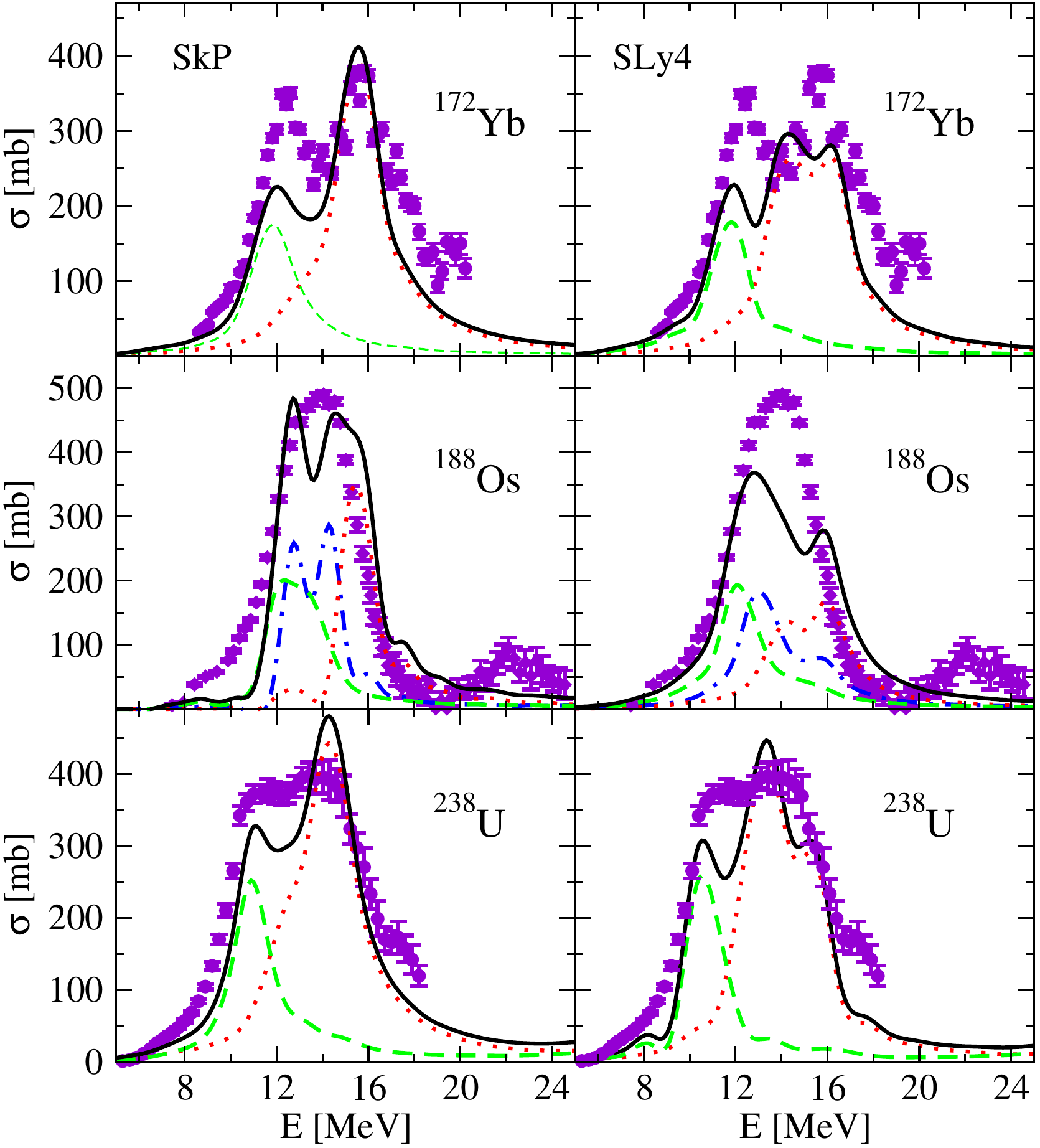}
\caption{ The calculated photo-absorption cross-section (solid black line), using two Skyrme force parametrizations for three deformed open-shell nuclei and the experimental $(\gamma,n)$ cross-sections (solid purple circles with error bars), extracted from Refence \cite{photo}. With dashed (green), dotted (red) and dotted-dashed (blue) lines we display the contribution to the cross-section arising from exciting the corresponding nucleus along the long axis, the short axis (multiplied by 2 for the prolate nuclei $^{172}$Yb and $^{238}$U) and the third middle axis in the case of the triaxial nucleus $^{188}$Os. \label{fig:14}}
\end{figure}

The emerging equations are formally equivalent to the TDHFB approximation with local potentials,  or to the time dependent Bogoliubov-de Gennes (TDBdG) equations:
\bea
i\hbar 
\left  ( \begin{array} {c}
  \dot{u}_{k\uparrow}\\ \dot{u}_{k\downarrow}\\ 
  \dot{v}_{k\uparrow} \\ \dot{v}_{k\downarrow}\\
\end{array} \right ) =
\left ( \begin{array}{cccc}
h_{\uparrow\uparrow}  & h_{\uparrow\downarrow} &0&\Delta \\
h_{\downarrow\uparrow}  & h_{\downarrow\downarrow} &-\Delta &0\\
0&-\Delta^* &-h^*_{\uparrow\uparrow} & -h^*_{\uparrow\downarrow}\\
\Delta^*&0&-h^*_{\uparrow\downarrow} & -h^*_{\downarrow\downarrow}
\end{array} \right )  
\left  ( \begin{array} {c}
  u_{k\uparrow}\\ u_{k\downarrow}\\ 
  v_{k\uparrow}\\ v_{k\downarrow}\\
\end{array} \right ). 
\eea
where we have suppressed the spatial ${\bf r}$ and time $t$ coordinates, and $k$ is the label of each qpwf  $[u_{k\sigma} ({\bf r},t), v_{k\sigma} ({\bf r},t)]$. where $\sigma=\uparrow,\downarrow$.  The single-particle Hamiltonian $h_{\sigma\sigma'}({\bf r},t)$ is a partial differential operator (thus local)  and $\Delta({\bf r},t)$ is a pairing field, all defined through the normal, anomalous, spin, and isospin densities and currents. The interaction with various applied external fields (spin, position and/or time-dependent) is described by including the corresponding potentials in the single-particle Hamiltonian $h_{\sigma\sigma'}({\bf r},t)$.  In one  of the first applications of TDSLDA to nuclei we have evaluated for the first time the photo-absorption cross-section on a trixially deformed open-shell nucleus $^{188}$Os. Within TDSLDA the nuclear response is described by applying an external electromagnetic field pulse and performing subsequently a Fourier analysis of the time response in order to extract the frequency dependent cross-section \cite{ionel}. Typically open shell nuclei display superfluidity in both proton and neutron systems. So far there are no indications, either experimentally or theoretically, that the two superfluids couple to each other in nuclei. The comparison between the theory and experiment shows a surprisingly good agreement without any fitting parameters, see Figure \ref{fig:14}, in spite of the known uncertainties of the nuclear energy density functional. It appears that so far the known nuclear energy density functional encapsulate reasonably well the gross nuclear properties.  

\section{Summary and outlook}

 The new framework TDSLDA, which is an extension of the DFT to the real-time dynamics of Fermi superfluids incorporates in a natural way the basic elements of the hydrodynamic approach at zero temperature (which is nothing else but a local implementation of conservation laws), but it also has the necessary quantum features to describe a wide range of topological excitations and moreover the Cooper pair breaking mechanism. The corresponding EDF is constructed using QMC input of homogeneous systems and the properties of the inhomogeneous systems are used to validate this functional, or further refine its properties as in Reference \cite{gradient}.  All required symmetries (translational, rotational, parity, gauge, Galilean, isospin, etc.), couplings to various external fields, and renormalizability of the theory are naturally incorporated into the theoretical framework.  Apart from being able to describe correctly known experimental facts, this new approach leads to new qualitative predictions (supercritical flow, quantum shock waves and domain walls, Higgs modes, vortex crossings, etc.) and it will allow as well to study the onset and dynamics of quantum turbulence in Fermi superfluids for the first time. It will be particularly interesting to perform further studies of the UFG under different conditions, in time-dependent optical lattices with or without varying the scattering length both in time and space, and in random external potentials, either static or dynamic. It is straightforward as well to couple the UFG within TDSLDA to artificial gauge fields \cite{agauge} and study its response. 
 
In nuclear physics studies TDSLDA appears as the only theoretical framework which would allow us to study on an equal footing both the structure and reactions of medium and heavy nuclei. Traditionally nuclear structure studies are performed using different theoretical approaches to describe nuclear structure and reactions, and this dichotomy can be put to rest within TDSLDA, in which an impinging neutron on a nucleus for example is described on a equal footing with the bound neutrons in the target nucleus. In this manner we hope to address for the first time the dynamics of induced fission with neutrons, gamma rays, and other projectiles.  Properties usually beyond reach within DFT, such as various two-body observable, surprisingly, can also be extracted using a method suggested quite some time ago by Balian and V\'en\'eroni \cite{balian1,balian2,balian3}, by implementing a very simple extension of TDSLDA \cite{long}. The dynamics of the vortex pinning and de-pinning in neutron star crust \cite{itoh}, and even the onset of turbulent motion due to vortex crossing and recombination \cite{link} are still unsettled issues. The vortex pinning mechanism in neutron star crust is still a matter of debate \cite{broglia1,broglia2,pierre} as it is not clear yet whether one has to perform static calculations of the vortex in the presence of nucleus at constant asymptotic density, constant neutron chemical potential or constant particle number in a simulation cell. When a vortex is formed a rather large mass redistribution occurs and the mass redistribution is further modified in the presence of a nucleus. The uncertainties arising from using various ensembles in conjunction with uncertainties in the nuclear EDF make the resolution of the question if a vortex is pinned or not on a nucleus rather unclear. Calculating the actual pinning force requires analyzing off-axis configurations that are not stationary. One can  characterize the vortex-nucleus dynamics through dynamical simulations, by forcing the nucleus to move with an external potential, the vortex (un)pinning can be characterized by looking at the rate of energy being transferred to the system \cite{sharma}.   
 
The author is not aware of any affiliations, memberships, funding, or financial holdings that might be perceived as affecting the objectivity of this review.

AB thanks G.F. Bertsch, J. Carlson, M.M. Forbes, Y.-L. Luo, P. Magierski, S. Reddy, K.J. Roche, I. Stetcu, S. Yoon, and Y. Yu for many discussions over the years. This work was supported by U.S. DOE Grants No. DE-FG02-97ER41014, No. DE-FC02-07ER41457 and the calculations have been performed on University of Washington Hyak cluster  (NSF MRI Grant No. PHY-0922770), Franklin (Cray XT4, NERSC, DOE Grant No. B-AC02-05CH11231), and JaguarPF (Cray XT5, NCCS, DOE Grant No. DE-AC05-00OR22725). 


\end{document}